\documentclass{aa}  

\usepackage{natbib}	
\usepackage{graphicx}
\usepackage{txfonts,epsfig,graphicx,url,twoopt}
\usepackage[breaklinks=true]{hyperref} 
\hypersetup{colorlinks=true,citecolor=blue,linkcolor=blue}
\bibpunct{(}{)}{;}{a}{}{,}     
\usepackage[usenames,dvipsnames]{color}
\DeclareMathOperator\erf{erf}

\begin{document}
   \title{Planetesimal formation via sweep-up growth at the inner edge \\ of dead zones}

   \author{J. Dr\k{a}\.{z}kowska\inst{1}\fnmsep\inst{2}
          \and
          F. Windmark\inst{1}\fnmsep\inst{2}
          \and
          C.P. Dullemond\inst{1}
          }

   \institute{Heidelberg University, Center for Astronomy, Institute for Theoretical Astrophysics, Albert-Ueberle-Str.\ 2, 69120 Heidelberg, Germany\\
         \email{drazkowska@uni-heidelberg.de}
         \and
         {Member of the International Max Planck Research School for Astronomy and Cosmic Physics at the Heidelberg University}
             }

   \date{Received 25 March 2013 \slash Accepted 13 June 2013}


 
  \abstract
   {
   The early stages of planet formation are still not well understood. Coagulation models have revealed numerous obstacles to the dust growth, such as the bouncing, fragmentation, and radial drift barriers. Gas drag causes rapid loss, and turbulence leads to generally destructive collisions between the dust aggregates.
   }
   {
   We study the interplay between dust coagulation and drift to determine the conditions in protoplanetary disk that support the formation of planetesimals. We focus on planetesimal formation via sweep-up and investigate whether it can take place in a realistic protoplanetary disk.
   }
   {
  We have developed a new numerical model that resolves the spatial distribution of dust in the radial and vertical dimensions.
  The model uses representative particles approach to follow the dust evolution in a protoplanetary disk. 
  The coagulation and fragmentation of solids is taken into account in the Monte Carlo method.
  A collision model adopting the mass transfer effect, which can occur for different-sized dust aggregate collisions, is implemented.
  We focus on a protoplanetary disk that includes a pressure bump caused by a steep decline of turbulent viscosity around the snow line. 
   } 
   {Our results show that high enough resolution of the vertical disk structure in dust coagulation codes is needed to obtain adequately short growth timescales, especially in the case of a low turbulence region.
We find that a sharp radial variation in the turbulence strength at the inner edge of dead zone promotes planetesimal formation in several ways. It provides a pressure bump that efficiently prevents the dust from drifting inwards. It also causes a radial variation in the size of aggregates at which growth barriers occur, favoring the growth of large aggregates by sweeping up of small particles. In our model, by employing an ad hoc $\alpha$ viscosity change near the snow line, it is possible to grow planetesimals by incremental growth on timescales of approximately 10$^5$ years.
}
   {}


   \maketitle

\section{Introduction}

Despite decades of research, planet formation is still not fully understood. At the point of formation, the protoplanetary disk is thought to contain submicron dust grains. The formation of planetesimals out of these grains is one of the more uncertain aspects in the theory of planet formation, since the growth of large dust particles by subsequent sticking collisions is very difficult to obtain in realistic models. Such a simple particle aggregation has been shown to encounter numerous obstacles, such as the electrostatic barrier \citep{2011ApJ...731...96O}, the bouncing barrier \citep{2010A&A...513A..57Z}, the fragmentation barrier \citep{1993Icar..106..151B}, and the radial drift barrier \citep{1977MNRAS.180...57W}. The relative velocities of dust particles, which are regulated by their interaction with gas, have been found to be too high to allow sticking of aggregates as small as millimeters. On the other hand, even if there is a way to grow meter-sized bodies, they are going to be lost inside of an evaporation line within a few hundred years due to the radial drift.

Some solutions to these problems have been suggested in recent years. 
The sticking properties of ices are claimed to be much better than those of silicates \citep{2009ApJ...702.1490W}, leading to ice grains capable of forming highly porous aggregates. Including this property in models has been shown to let the particles avoid the radial drift barrier \citep{2012ApJ...752..106O}. However, the collisional properties of the ice particles are still rather uncertain, because of the difficulties in conducting the laboratory experiments. There is much more laboratory data concerning the collisional physics of the silicates \citep{2010A&A...513A..56G}, although even the silicate collisional properties remain an extensively discussed topic. Recent experiments have revealed a smooth transition between sticking and bouncing behavior \citep{2008ApJ...675..764L, 2012Icar..218..688W, 2013arXiv1302.5532K}, but numerical molecular dynamics simulations predict no or significantly less bouncing \citep{2011ApJ...737...36W, 2013A&A...551A..65S} At still higher collision velocities, particles are expected to fragment, but if the mass ratio is high enough, growth via mass transfer is also a possibility \citep{2005Icar..178..253W, 2009MNRAS.393.1584T, 2010ApJ...725.1242K, 2011ApJ...736...34B}.

The relative velocity of collision between the aggregates is usually calculated on the basis of a mean turbulence model, where two particles with given masses $m_1$ and $m_2$ always collide at the same relative velocity $\Delta v(m_1,m_2)$. Considering a probability distribution function $P(\Delta v|m_1,m_2)$ and the sweep-up by the mass transfer is another possibility of overcoming the growth barriers \citep{2012A&A...544L..16W, 2013ApJ...764..146G}. However, as the exact nature of the probability distribution is unknown, it is not certain if this effect can indeed allow the planetesimal growth.
The combined action of hydrodynamic and gravitational instabilities \citep{2000Icar..148..537G,2007Natur.448.1022J} is an alternative scenario for the formation of planetesimals. 
The radial drift barrier can be overcome by taking local disk inhomogeneities into account that lead to the pressure gradient change (pressure bumps) and suppress of the inward drift of bodies \citep{1972fpp..conf..211W, 1995A&A...295L...1B, 1997Icar..128..213K, 2007MNRAS.375..500A, 2007ApJ...671.2091G, 2007ApJ...664L..55K}.

Modeling the planet formation is not only difficult because of the growth barriers at the first stage of the process.
Formation of a single planet covers about 40 orders of magnitude in mass, which is not possible to handle with any traditional method, because of a fundamental difference in the physics involved in its different stages. In the small particle regime, there are too many independent particles for an individual treatment. The coagulation is driven by random collisions. Therefore, statistical methods are used to model the evolution of the fine dust medium \citep{1980Icar...44..172W,1981Icar...45..517N,2008A&A...480..859B,2010A&A...513A..79B}. In this approach, the dust medium is followed using the grain distribution function $f_{\rm{d}}(m,r,t)$, giving the number of dust particles of particular properties at a given time. In the big body regime, the evolution is led by gravitational dynamics. That forces us to treat the objects individually using N-body methods \citep{2000Icar..143...15K}. A connection between the two methods requires an ad hoc switch. Such a solution has been implemented by \citet{1991Icar...92..147S}, \citet{2006AJ....131.2737B} and recently \citet{2011arXiv1105.6094G}.

In addition to the statistical methods mentioned above, there are also Monte Carlo methods used in the small particle regime \citep{1975MNRAS.170..541G, 2007A&A...461..215O}.
In recent years, a new kind of algorithm has been developed: a Monte Carlo algorithm with the representative particle approach \citep{2008A&A...489..931Z}.
In this method, the huge number of small particles is handled by grouping the (nearly) identical bodies into swarms and representing each swarm by a representative particle. Instead of evolving the distribution function $f_{\rm{d}}(m,r,t)$, it is sampling and reproducing it with the use of the representative bodies. This manner should allow a much smoother and more natural transition to the N-body regime.
Indeed, this kind of approach is already used in the N-body codes. \citet{2010AJ....139.1297L} applied a superparticle approach to treat planetesimals. They showed that taking the gravitational interactions into account is very important in the case of kilometer size bodies. The gravitational interplay can lead to redistribution of the material and change accretion rates in the protoplanetary disk.

With the work presented in this paper, we make the very first step toward a new computational model that will connect the small scale dust growth to the large scale planet formation. We develop a 2D Monte Carlo dust evolution code accounting both for drift and coagulation of the dust particles. We expect to extend this method in future by adding the gravitational interactions between the bodies.

This paper is organized as follows. We introduce our numerical model in Sect.\ \ref{sub:model}. We demonstrate some tests of the code in Sect.\ \ref{sub:tests}. In Sect.\ \ref{sub:1D}, we show results obtained with the 1D version of our code and compare them with results presented by \citet{2011A&A...534A..73Z}. In Sect.\ \ref{sub:2D}, we present results obtained with the 2D version of the code, showing that applying a disk model including a steep variation of turbulent strength near the snow line \citep{2007ApJ...664L..55K} and a collisional model that takes the mass transfer effect into account \citep{2012A&A...540A..73W}, allows us to overcome the bouncing barrier and turn a limited number of particles into planetesimals on the timescale of approximately 10$^5$ yrs. We provide discussion of the presented results as well as conclusions in Sect.~\ref{sub:last}.

\section{The numerical model}\label{sub:model}
We develop a 2D Monte Carlo dust evolution code, able to resolve a protoplanetary disk structure in radial and vertical dimension. We assume that the disk is cylindrically symmetric, thereby ignoring the azimuthal dependence. We use an analytical description for the gas disk. The dust is treated using the representative particle approach. The code is a further development of the work presented by \citet{2011A&A...534A..73Z}. The code is written in Fortran 90 and is parallelized using OpenMP directives.

In each time step the code performs the following steps:
\begin{enumerate}
\item{Advection velocities of the dust particles are determined taking their current properties and positions into account.}
\item{The code time step is calculated on the basis on the advection velocities and existing grid, following the Courant condition.}
\item{Advection of the particles is performed both in radial and vertical direction. The solids undergo the radial drift, vertical settling and turbulent diffusion.}
\item{The new grid is established according to the updated positions of the particles, using the adaptive grid routine (see Fig.~\ref{fig:ag}).}
\item{Collisions between the particles are performed in each cell by the Monte Carlo algorithm. The particle properties are updated.}
\item{The output is saved when required.}
\end{enumerate}
More detailed description of the approach used in the code can be found in the following sections. 

\subsection{Gas description}
The gas structure is implemented in the form of analytical expressions for the gas surface and volume density $\Sigma_{\rm{g}}(r)$ and $\rho_{\rm{g}}(r,z)$, pressure $P_{\rm{g}}(r)$, temperature $T_{\rm{g}}(r)$ and turbulent viscosity $D_{\rm{g}}(r)$.

For now we assume that the gas in the disk does not evolve, although this is not a fundamental restriction, and the gas evolution is possible to implement without severe changes in the code structure. In a first-order approximation, the time evolution can be implemented analytically by expanding the gas properties description from the function of space $f_{\rm{g}}(r)$ to the function of space and time $f_{\rm{g}}(r,t)$.

\subsection{Dust description}\label{sub:dustd}
To describe the dust, we use the approach based on \citet{2008A&A...489..931Z}. We follow the “lives” of $n$ representative particles, which are supposed to be a statistical representation of $N$ physical particles present in an examined domain. Commonly $n \ll N$. For a typical protoplanetary disk, with mass of 0.01~$M_{\odot}$ and a dust to gas ratio of 0.01, consisting of 1 $\mu$m size dust grains, we would have $N\approx10^{42}$. For computational feasibility we would have e.g. $n=10^5$, meaning each representative particle $i$ represents $N_i\approx10^{37}$ physical particles.

All of the $N_i$ physical particles, represented by a single representative particle $i$, share identical properties: for now these are mass $m_i$ and location in the disk $(r_i,z_i)$. As we impose the axial symmetry, we do not include the azimuthal position. We assume that the physical particles belonging to one swarm are homogeneously distributed along an annulus of given location $r_i$ and height above the midplane $z_i$. The total mass of physical particles contained in one swarm $M_{\rm{swarm}}=N_im_i$ is identical for every representative particle and it does not change with time. This means that the $N_i$ has to drop when the particle mass $m_i$ grows. This is not a physical effect, just a statistical. See \citet{2008A&A...489..931Z} for details.

With the representative particle approach, it is relatively easy to add further dust properties, in particular the internal structure of aggregates, which was shown to be important by \citet{2007A&A...461..215O}. We leave the implementation of the porosity for further work.

When performing the advection, we assume that all of the physical particles in the swarm undergo the same change of the position $(r_i,z_i)$ and after the shift, they are still uniformly distributed along the designated annulus. However, when we consider the collisions, we set up a numerical grid in order to account for the fact that only particles that are physically close can collide. In this case, we assume that the particles are homogeneously distributed inside a grid cell (see Sect.\ \ref{sub:coll} for description of the grid). This assumption is required by the method used to investigate the collisional evolution of the aggregates \citep{2008A&A...489..931Z}. The difference between the locations assumed in both of the cases is most often not important and can be treated as a kind of systematic uncertainty. 

\subsection{Advection of dust particles} 
The location of a representative particle changes because of radial drift and vertical settling as well as turbulent diffusion. The main particle characteristics determining its behavior with respect to gas is so called Stokes number $\rm{St}$. 
It is defined as
\begin{equation}\label{stokes}
{\rm{St}} = t_{\rm{s}} \Omega_{\rm{K}},
\end{equation}
where $\Omega_{\rm{K}}$ denotes the Kepler frequency and $t_{\rm{s}}$ is the so-called stopping time of the particle.
The Stokes number can be treated as a particle-gas coupling strength indicator. If $\rm{St} \ll 1$, the particle is well coupled to the ambient gas and its motion is fully dependent on the motion of the gas. On the other hand, the particles with $\rm{St} \gg 1$ are practically independent of the gas.

The stopping time of the particle $t_{\rm{s}}$ determines a timescale that the particle needs to adjust its velocity to the velocity of the surrounding gas. 
The exact expression that we use to compute the $t_{\rm{s}}$ depends on the particle radius $a$. The ratio of the mean free path of the gas $\lambda_{\rm{mfp}}$ and the particle size $a$ is called Knudsen number~$\rm{Kn}$:
\begin{equation}\label{Kn}
{\rm{Kn}} = \frac{\lambda_{\rm mfp}} {a}.
\end{equation}
If a particle's Knudsen number is $\rm{Kn} > 4\slash9$, the particle is in the Epstein drag regime and its stopping time is given by \citep{1977MNRAS.180...57W} 
\begin{equation}\label{stEp}
t_{\rm{s}}^{\rm{Ep}} = \frac{a \rho_{\rm{p}}} { v_{\rm{th}} \rho_{\rm{g}}},
\end{equation}
where $\rho_{\rm{p}}$ is the internal density of the particle and $v_{\rm{th}}$ is the thermal velocity of the gas. The latter is expressed as  $v_{\rm{th}}=\sqrt{8 k_{\rm{B}} T_{\rm{g}} \slash \pi m_{\rm{g}}}$, where $k_{\rm{B}}$ is the Boltzmann constant, $T_{\rm{g}}$ is the gas temperature and $m_{\rm{g}}$ is mass of the gas molecule. 
On the other hand, when $\rm{Kn} < 4\slash9$, the particle is in the Stokes regime. The Stokes regime is in general not homogeneous and is often divided into subregimes. The Reynolds number of the particles $\rm{Re}_{\rm{p}}$ defines which of the subregimes applies \citep{1977MNRAS.180...57W}. The $\rm{Re}_{\rm{p}}$ is specified as
\begin{equation}\label{Re_p}
{\rm{Re}}_{\rm{p}} = \frac{2a\Delta v_{\rm{pg}}}{\nu_{\rm{g}}},
\end{equation}
with $a$ denoting the particle radius, $\Delta v_{\rm{pg}}$ the relative velocity between the particle and the gas and $\nu_{\rm{g}}$ the molecular viscosity of gas that is expressed as $\nu_{\rm{g}}=v_{\rm{th}}\lambda_{\rm{mfp}}\slash2$.
As long as $\rm{Re}_{\rm{p}} < 1$, the first Stokes regime applies. In our models $\rm{Re}_{\rm{p}} > 1$ translates into $a \gtrsim 10^4$ cm. This is larger than we obtain in the models presented in this paper. Thus, for now we do not include the other Stokes regimes. For the particles with $\rm{Kn} < 4\slash9$ we assume \citep{1977MNRAS.180...57W}
\begin{equation}\label{stSt}
t_{\rm{s}}^{\rm{St}} = t_{\rm{s}}^{\rm{Ep}} \times \frac{4}{9}{\rm Kn}^{-1}.
\end{equation}

\paragraph{Radial drift}
The radial drift of dust particles has two sources. One of them is the gas accretion onto the central star. The gas moves inwards and drags the dust particles with it. This phenomenon is stronger for small particles ($\rm{St} \ll 1$), and it is not important for big ones ($\rm{St} \gg 1$). The drift velocity caused by this effect can be expressed as \citep{2008A&A...480..859B}
 \begin{equation}\label{vdrift1}
v_{\rm{d}}^{\rm{acc}} = \frac{v_{\rm{g}}^{\rm{r}}}{1+{\rm{St}}^2},
\end{equation}
where $v_{\rm{g}}^{\rm{r}}$ denotes the accretion velocity of gas. We use a convention in which the $v_{\rm{g}}^{\rm{r}} < 0$ indicates inward drift.

The other effect is also related to the coupling of the solids to gas, but now the radial drift is a result of orbital movement. In a gas-free environment, the solid particles orbit around the star with the Keplerian velocity $v_{\rm{K}}$, resulting from a balance between the gravity and the centrifugal force. For the gas however, the pressure needs to be considered. Therefore, the gas moves with a sub-Keplerian velocity. Because of the difference in the azimuthal velocity of gas and dust, the dust particles feel a constant head-wind. Interacting with the gas via the drag force, they loose the angular momentum and thus drift inwards with velocity \citep{1977MNRAS.180...57W,2008A&A...480..859B}:
\begin{equation}\label{vdrift2}
v_{\rm{d}}^{\rm{drift}} = \frac{2v_{\eta}}{\rm{St}+\frac{1}{St}}.
\end{equation}
Hence, this effect is not significant for both very small and very big dust particles, but for the particles with $\rm{St}\approx1$ the drift velocity $v_{\rm{d}}^{\rm{drift}}$ can reach even $30$ m s$^{-1}$ \citep{2008A&A...480..859B}.

The maximum drift velocity $v_{\eta}$ can be expressed as \citep{2008A&A...487L...1B}
\begin{equation}\label{veta}
v_{\eta} = \frac{\partial_{\rm{r}} P_{\rm{g}}}{2 \rho_{\rm{g}} \Omega_{\rm{K}}}.
\end{equation}
The $v_{\eta}$ is dependent on the gas pressure gradient $\partial_{\rm{r}} P_{\rm{g}}$, which can be both negative (in most of the standard disk models it is negative over the whole disk) and positive. If we find a disk model, in which locally $\partial_{\rm{r}} P_{\rm{g}} > 0$ (a so-called {\it pressure bump}), we get outward radial drift of solids that leads to a local significant enhancement of dust density. 

The total radial drift velocity is given by 
\begin{equation}\label{vdrifttot}
v_{\rm{d}}^{\rm{r}} = v_{\rm{d}}^{\rm{acc}} + v_{\rm{d}}^{\rm{drift}}.
\end{equation}

\paragraph{Vertical settling}
The dust particles present in the protoplanetary disk are settling down towards the midplane due to gravity from the central star. The settling velocity is regulated by the gas drag. It can be obtained from basic equations as \citep{2004A&A...421.1075D}
\begin{equation}\label{vsvel}
   v_{\rm{d}}^{\rm{z}} = -z \Omega^2_{\rm{K}} t_{\rm{s}},
\end{equation}
where $z$ is the height above the midplane. It can be rewritten using Eq.\ (\ref{stokes}) as
\begin{equation}\label{vsvel2}
   v_{\rm{d}}^{\rm{z}} = -z \Omega_{\rm{K}} \rm{St}.
\end{equation}
For big particles, the velocity calculated from Eq.\ (\ref{vsvel2}) would be higher than the orbital velocity projected on the $z$ axis, so we restrict it to
\begin{equation}\label{vsvelmin}
   v_{\rm{d}}^{\rm{z}} = -z \Omega_{\rm{K}} \min(0.5,\rm{St}),
\end{equation}
following e.g. \citet{2010A&A...513A..79B}.
This description is not valid for big particles that are completely decoupled from the gas. These particles undergo the orbital oscillations around the midplane. A direct integration of the equations of motion would need to be included in order to account for this effect. We leave it for further work.

\paragraph{Turbulent diffusion}
If there were no other effects in the disk, all the dust would eventually form an infinitely thin layer in the midplane. However, we assume that there is a turbulence present in the disk. We implement the effect of the turbulence on the particles spatial distribution in the same way to \citet{2010ApJ...723..514C} and \citet{2011A&A...534A..73Z}. We take the diffusion in both vertical and radial directions into account. 

The turbulence generally smears out the density distribution (turbulent diffusion). If we take a point dust distribution after time $t$ it will become a Gaussian distribution with the half width $L$ (in 1D):
\begin{equation}\label{sigma}
L = L(t) = \sqrt{2D_{\rm{d}}t},
\end{equation}
where $D_{\rm{d}}$ is the dust turbulent diffusion coefficient, which we can express as
\begin{equation}\label{ddust}
   D_{\rm{d}} = \frac{D_{\rm{g}}}{\rm{Sc}},
\end{equation}
where $\rm{Sc}$ is called the Schmidt number, and the gas diffusion coefficient $D_{\rm{g}}$ (turbulent viscosity) is assumed to have the form of so-called $\alpha$ viscosity \citep{1973A&A....24..337S}:
\begin{equation}\label{dgas}
   D_{\rm{g}} = \alpha c_{\rm{s}} H_{\rm{g}}.
\end{equation}
$\alpha$ is a parameter describing the efficiency of the angular momentum transport with values typically much lower than $1$. $c_s$ is the sound speed in gas and $H_{\rm{g}}$ is gas pressure scale height, which is expressed as $H_{\rm{g}} = {c_{\rm{s}}}\slash{\Omega_{\rm{K}}}$. The Schmidt number is currently estimated as \citep{2007Icar..192..588Y,2011MNRAS.415...93C}
\begin{equation}\label{schmidt}
   \rm{Sc} = 1 + St^2.
\end{equation}
We implement the turbulent diffusion of the solid particles as random jumps. We add a term corresponding to our turbulence prescription to the velocity resulting from the radial drift and vertical settling. The turbulent velocity resulting from the prescription given above is
\begin{equation}\label{vD1}
   v_{\rm{d}}^{\rm{D1}} = \frac{\Delta x}{\Delta t},
\end{equation} 
where $\Delta x$ is the displacement of the particle during the time step $\Delta t$. The displacement is taken as a random number drawn from a Gaussian distribution with the half width $L$ from Eq.\ (\ref{sigma}). 

This description of the diffusion is however simplified. In fact, there is an additional term in the diffusion equation for a non-homogeneous gas distribution. The velocity component resulting from this effect always points towards the gas density maximum. Therefore, the dust scale height never exceeds the gas scale height. For more details see \citet{2011A&A...534A..73Z} (their Eqs 7-8). The velocity corresponding to this term can be noted as
\begin{equation}\label{vD2}
   v_{\rm{d}}^{\rm{D2}} = D_{\rm{d}}\frac{1}{\rho_{\rm{g}}}\frac{\partial\rho_{\rm{g}}}{\partial x},
\end{equation}
where $x$ in Eqs (\ref{vD1}) and (\ref{vD2}) can be both $r$ and $z$, depending on direction along that we consider the diffusion.

\subsection{Collisions}\label{sub:coll}
\paragraph{Monte Carlo method}
We model the dust coagulation using a Monte Carlo algorithm. This approach was already used in the protoplanetary disk context by \citet{2007A&A...461..215O}. It is based on a method presented for the first time by \citet{1975MNRAS.170..541G}. \citet{2008A&A...489..931Z} described in detail how to use this algorithm with the representative particles approach. Only the main facts are stated here for the reader's convenience. 

As mentioned in Sect.\ \ref{sub:dustd}, we assume that a limited number $n$ representative particles represent all $N$ physical particles present in the computational domain. Each representative particle $i$ describes a swarm of $N_i$ identical physical particles. Total mass $M_{\rm{swarm}}$ of every swarm is equal and constant in time. 

As we typically have $n \ll N$, we only need to consider the collisions between representative and non-representative particles. The collisions between the representative particles are too rare to be significant. The collisions among the physical particles do not need to be tracked as the basic assumption of the method. 

The particles taking part in the subsequent collisions as well as the time step between the events are determined on a basis of random numbers. For each collision we pick one representative particle $i$ and one non-representative particle from the swarm represented by the representative particle $k$. It is possible that $i=k$. The probability of a collision between particles $i$ and $k$ is determined as
 \begin{equation}\label{rik}
   r_{ik} = \frac{N_k K_{ik}}{V},
\end{equation}
where $V$ is the cell volume and $K_{ik}$ is a coagulation kernel. Apart from some test cases we use
\begin{equation}\label{Kik}
   K_{ik} = \Delta v_{ik} \sigma_{ik},
\end{equation}
where $\Delta v_{ik}$ is the relative velocity between particles $i$ and $k$ and $\sigma_{ik}$ is the geometrical cross-section for their collision.
The total collision rate among any of the pairs is
\begin{equation}\label{totr}
   r = \sum\limits_i\sum\limits_k r_{ik}.
\end{equation}
We first choose the representative particle, and the probability that it is particle $i$ is
\begin{equation}\label{Pi}
   P_{i} = \frac{\sum\limits_k r_{ik}}{r}.
\end{equation}
Then we choose the non-representative particle with the probability:
\begin{equation}\label{Pik}
   P_{k|i} = \frac{r_{ik}}{\sum\limits_k r_{ik}}.
\end{equation}
The time step between the subsequent collisions is determined as 
\begin{equation}\label{timestepMC}
   \tau = -\frac{1}{r}\ln({\mbox{rand}}),
\end{equation}
where the $\mbox{rand}$ is a random number drawn from the uniform distribution between 0 and 1.

As a result of the collision, only the representative particle $i$ changes its properties. For example, in the case of sticking, $m_i~\leftarrow~m_i~+~m_k$. Every time the mass of the particle changes, the number of particles represented by the swarm has to be updated as $N_i = M_{\rm{swarm}} \slash m_i$.

\begin{figure}
   \centering
   \includegraphics[width=\hsize]{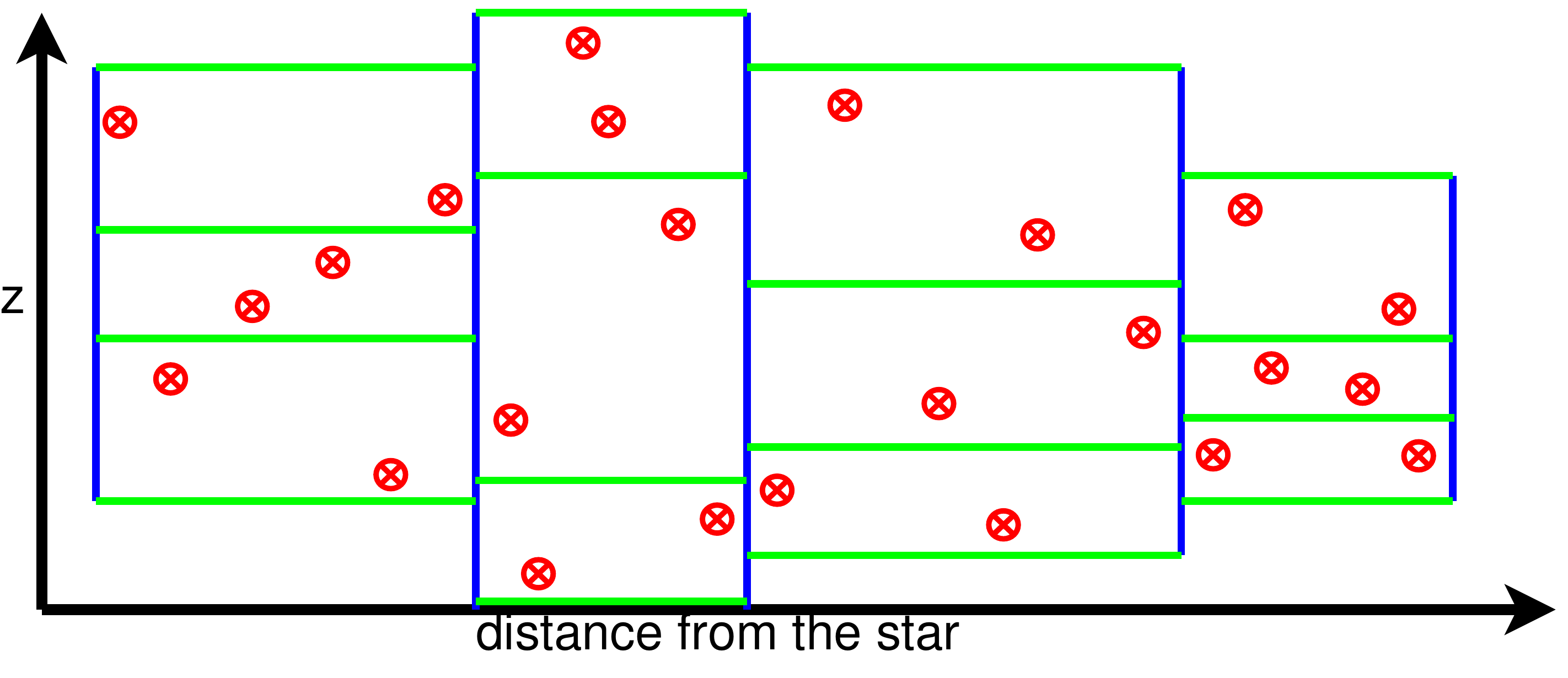}
      \caption{Illustration of the adaptive grid algorithm. The dots correspond to the representative particles. First the vertical (blue in color version) walls are established so that the number of the representative particles in each radial zone is equal. Then the horizontal (green) walls are set up for each radial zone individually in order to preserve equal number of swarms in each cell.}
         \label{fig:ag}
\end{figure}   
\paragraph{Adaptive grid}
The coagulation of dust aggregates depends on the local properties of the ambient gas. This is the reason why, to perform the collisions, we first set up a 2D ($r+z$) grid and place our representative particles in the grid cells. The grid cells are assumed to be annuli at a given distance from the star $\left\{r,r+\Delta r\right\}$ and height above the midplane $\left\{z,z+\Delta z\right\}$. Only particles present inside the same annulus are allowed to collide with each other.

To construct the annuli we developed an adaptive grid routine. The volume of the grid cells varies in order to keep the number of the swarms per cell constant. This procedure is illustrated in Fig.\ \ref{fig:ag}. In order to set up the grid walls, we first sort the particles by their radial positions. We choose the positions of the vertical walls such that the number of swarms in each radial zone is the same. Then we sort the particles by their vertical positions within every radial zone individually and set up the horizontal walls in order to preserve equal number of swarms in each cell.

Thanks to this approach, we automatically gain higher spatial resolution in the important high dust density regions. Furthermore, keeping the number of the representative particles in one cell constant assures that we always have a sufficient amount of bodies to resolve the physics of the coagulation kernel properly (see Sect.~\ref{sub:test1}).

As the Monte Carlo algorithm is generally an $O(n^2)$ method, the adaptive grid routine helps us to optimize the computational cost of performing the collisions by a significant factor.

\paragraph{Relative velocities}
As in e.g. \citet{2010A&A...513A..79B}, we consider five sources of relative velocities between the dust particles: namely the Brownian motion, turbulence, radial and azimuthal drift as well as differential settling. For the turbulent relative velocities we follow the prescription given by \citet{2007A&A...466..413O}.

For calculation of the relative velocities, all the particles are assumed to be in the center of the cell. Due to this, we avoid unphysically high collision velocities that could occur e.g. in case of a big cell with one particle placed on significantly higher height above the midplane $z$ than the other one. In such a situation, the relative velocity calculated on a basis of Eq.\ (\ref{vsvelmin}) is dominated by the difference of the height $z$. In reality, at the moment of the collision, $z$ is identical for both particles and the relative velocity is set up by the difference of the Stokes numbers. 

\subsection{Time step}
In order to resolve both advection and coagulation of the dust particles properly, a limit to the time step of the code is required. A drifting particle should be allowed to interact with every other particle along its way, thus it cannot “jump over” any cell. We implement an adaptive time-stepping method. We limit the time step according to the Courant condition:
\begin{equation}\label{courant}
   \Delta t^x < \frac{\Delta x_{\rm{min}}}{v_{\rm{max}}^{x}},
\end{equation}
where $x$ can be both $r$ and $z$, as we apply this condition to both directions and we finally choose $\Delta t =\min(\Delta t^{\rm{r}},\Delta t^{\rm{z}})$. $\Delta x_{\rm{min}}$ is the length of the shortest cell in the given direction and $v_{\rm{max}}^x$ is the drift velocity of the fastest particle in this direction. 
The final time step we obtain is typically of the order of a fraction of the local orbital period.

Generally, the time step should be limited also by the dust growth timescale. However, in typical cases, the advection timescale is shorter than the growth timescale. This means that within one advection time step, the coagulation does not change the drift properties significantly.

\begin{figure*}
   \centering
   \includegraphics[width=\hsize]{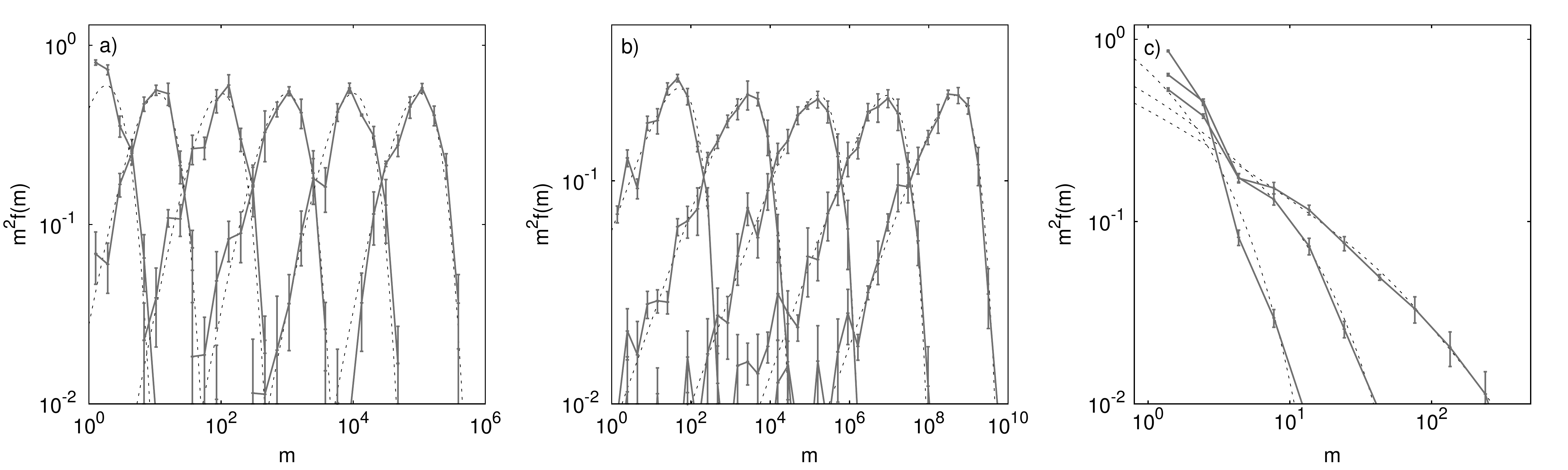}
   \caption{The grains mass distribution for the tests against the analytical solutions of the Smoluchowski equations (dashed lines) at different time instants: a) Test against the constant kernel $K_{ik}=1$, where $50$ representative particles are simulated five times. The particles masses are binned and the distribution functions are averaged at dimensionless times $t = 1,10,10^2,10^3,10^4,10^5$. b) Test against the linear kernel $K_{ik}=\frac{1}{2}\left(m_i+m_k\right)$. There are $150$ particles used and the simulation is repeated five times. The distribution function is produced at times $t = 4,8,12,16,20$. c) Test against the product kernel $K_{ik}=m_i \times m_k$. We use $400$ representative particles and repeat the simulation ten times. The outputs are produced at times $t = 0.4,0.7,0.9$.}
              \label{fig:mctests}%
    \end{figure*}
\section{Test cases}\label{sub:tests}
In order to validate our code, we perform a set of different tests. We test the advectional and collisional parts of the code separately as well as both of them together. In this section we present some of the more educative test results.

\subsection{Tests of the coagulation model}\label{sub:test1}
We test our implementation of the Monte Carlo coagulation method with the representative particle approach. In a 0-dimensional case, the only property of particles is their mass. In such case, the coagulation can be described by the Smoluchowski equation \citep{1916ZPhy...17..557S}. For some coagulation kernels $K_{ik}$, one can find analytical solution of the Smoluchowski equation. We test our approach against three such kernels, namely the constant kernel $K_{ik} = 1$, the linear kernel $K_{ik} = \frac{1}{2}(m_i + m_k)$ and the product kernel $K_{ik} = m_i \times m_k$. The tests results are presented in Fig.\ \ref{fig:mctests}. We start all the simulations with a homogeneous mass distribution of particles with $m_0=1$. The volume density of particles is also equal to unity. The analytical solutions are obtained from \citet{1979ApJ...229..242S} and \citet{1990Icar...88..336W}.

We necessarily get similar results as \citet{2008A&A...489..931Z}. We find that the constant kernel can be properly resolved using a very limited number of representative particles. The linear kernel is possible to resolve using at least 100 representative particles. To obtain proper evolution in the case of the product kernel we need about 300 particles. As the mass dependence of the coagulation kernel in physical cases usually lies between the linear and product kernels, we conclude that we should use at least 200 representative particles per cell in our simulations. Thanks to the adaptive grid routine, it is possible to fulfill this requirement at any time during the simulation.

\subsection{Vertical settling and turbulent diffusion of the particles}\label{sub:vert}
In the case of absence of the radial drift and coagulation, the vertical structure of dust is modulated by the vertical settling and turbulent diffusion. From the test simulations, we obtain a Gaussian distribution defined by local properties of gas and solids. Its width can be derived comparing the timescales of the vertical settling and turbulent diffusion.

The timescale of the vertical settling can be obtained from Eq.\ (\ref{vsvelmin}) as
\begin{equation}\label{vstime}
   \tau_{\rm{sett}} \approx \frac{1}{\Omega_{\rm{K}} \min(0.5,\rm{St})},
\end{equation}
The timescale of the turbulent diffusion can be estimated as
\begin{equation}\label{difftime}
   \tau_{\rm{diff}} \approx \frac{L^2}{D_{\rm{d}}}, 
\end{equation}
where the $L$ is a length scale over which the diffusion takes place and the $D_{\rm{d}}$ is defined by Eq.\ (\ref{ddust}). 
Comparing Eqs (\ref{difftime}) and (\ref{vstime}) and transforming the resulting formula using Eqs (\ref{ddust}) - (\ref{schmidt}) and taking $L = h_{\rm{d,1}}$ we can estimate the thickness of the dust layer as
\begin{equation}\label{hdust1}
   h_{\rm{d,1}} = H_{\rm{g}} \left( \frac{\alpha}{\min(0.5,{\rm{St}}) (1 + {\rm{St}}^2)} \right)^{1\slash2}.
\end{equation}
The above estimate does not take the part of diffusion introduced with Eq.\ (\ref{vD2}) into account. This effect prevents the dust layer scale height from exceeding the gas scale height. An analytical solution of the advection-diffusion equation of the gas disk gives a more accurate expression for the height of the dust layer \citep{1995Icar..114..237D}:
\begin{equation}\label{hdust}
  h_{\rm{d}} = h_{\rm{d,1}} \left[ 1+ \left(\frac{h_{\rm{d,1}}}{H_{\rm{g}}}\right)^2 \right]^{-1\slash2}.
\end{equation}

\begin{figure}
   \centering
   \includegraphics[width=\hsize]{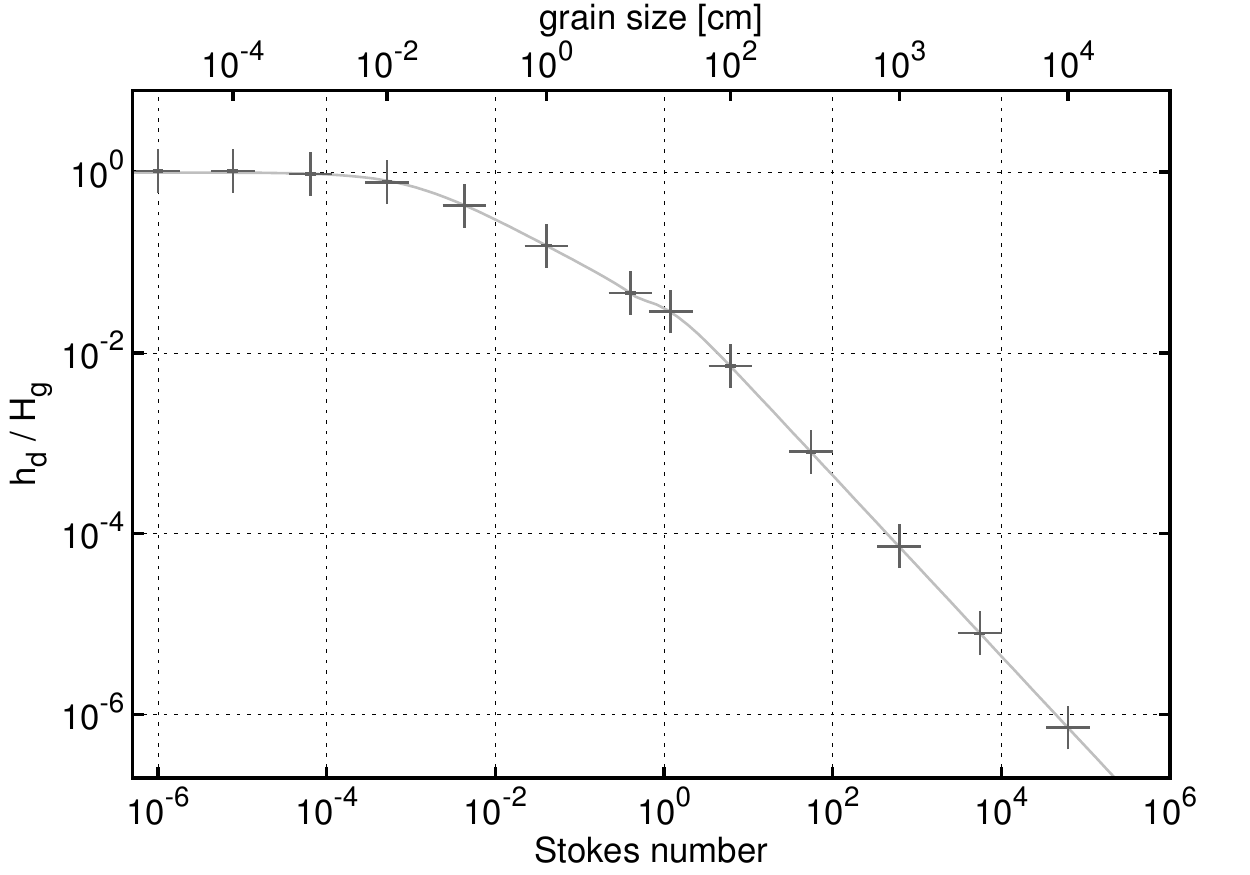}
      \caption{The results of the vertical settling and turbulent diffusion test. The theoretical dependence given by Eq.\ (\ref{hdust}) is plotted with the solid line. The change of the slope around $\rm{St}=0.5$ comes from the Stokes number restriction applied in Eq.\ (\ref{vsvelmin}). The test simulations results are marked with points. 
We find a good agreement between the analytical prediction and the tests results.
              }
         \label{fig:verticaltest}
\end{figure}

We perform a set of test runs to check if the dependence given by Eq.\ (\ref{hdust}) is reproduced by our code. We place the representative particles in a local column of a disk around a star with mass $M_\star = M_\odot$. The column is located at $r=1$ AU and we assume that a gas surface density $\Sigma_{\rm{g}}=100$ g cm$^{-2}$, temperature $T_{\rm{g}}=200$ K, and $\alpha=10^{-3}$ at this location. The initial dust to gas ratio is taken to be 0.01 and the dust material density 1.6 g cm$^{-3}$.
The gas vertical distribution is assumed to be Gaussian with the standard deviation of $H_{\rm{g}}$. Initially we place the representative particles such that we get constant dust to gas ratio at every height above the midplane, so $h_{\rm{d},0}=H_{\rm{g}}$. We use particles with sizes ranging from 10$^{-5}$ to 10$^4$ cm, corresponding to the Stokes numbers range of 10$^{-6}$ to 10$^5$. The radial drift and collisions are switched off for this test. After the particle distribution reaches a steady state, we measure $h_{\rm{d}}$ by fitting a Gaussian. Results of the test are presented in Fig.\ \ref{fig:verticaltest}. For each of the runs we use 10$^4$ representative particles distributed over 100 cells. We find a good agreement between the analytical prediction (Eq.\ \ref{hdust}) and the test results. 

\subsection{Trapping of the dust particles in a pressure bump}\label{sub:trap}
The trapping of solids in a region with positive pressure gradient is a promising mechanism of overcoming the radial drift barrier and enhancing the growth of dust aggregates \citep{2007ApJ...664L..55K,2008A&A...487L...1B}. It was already studied theoretically by e.g. \citet{2007ApJ...671.2091G}. \citet{2012A&A...538A.114P} investigated trapping of solids in the outer regions of protoplanetary disk. They showed that disk models with pressure bumps give predicted spectral slope in the mm-wavelength range consistent with the observed for typical T-Tauri disks, contrary to disk models without the bumps.

\begin{figure}
   \centering
   \includegraphics[width=0.9\hsize]{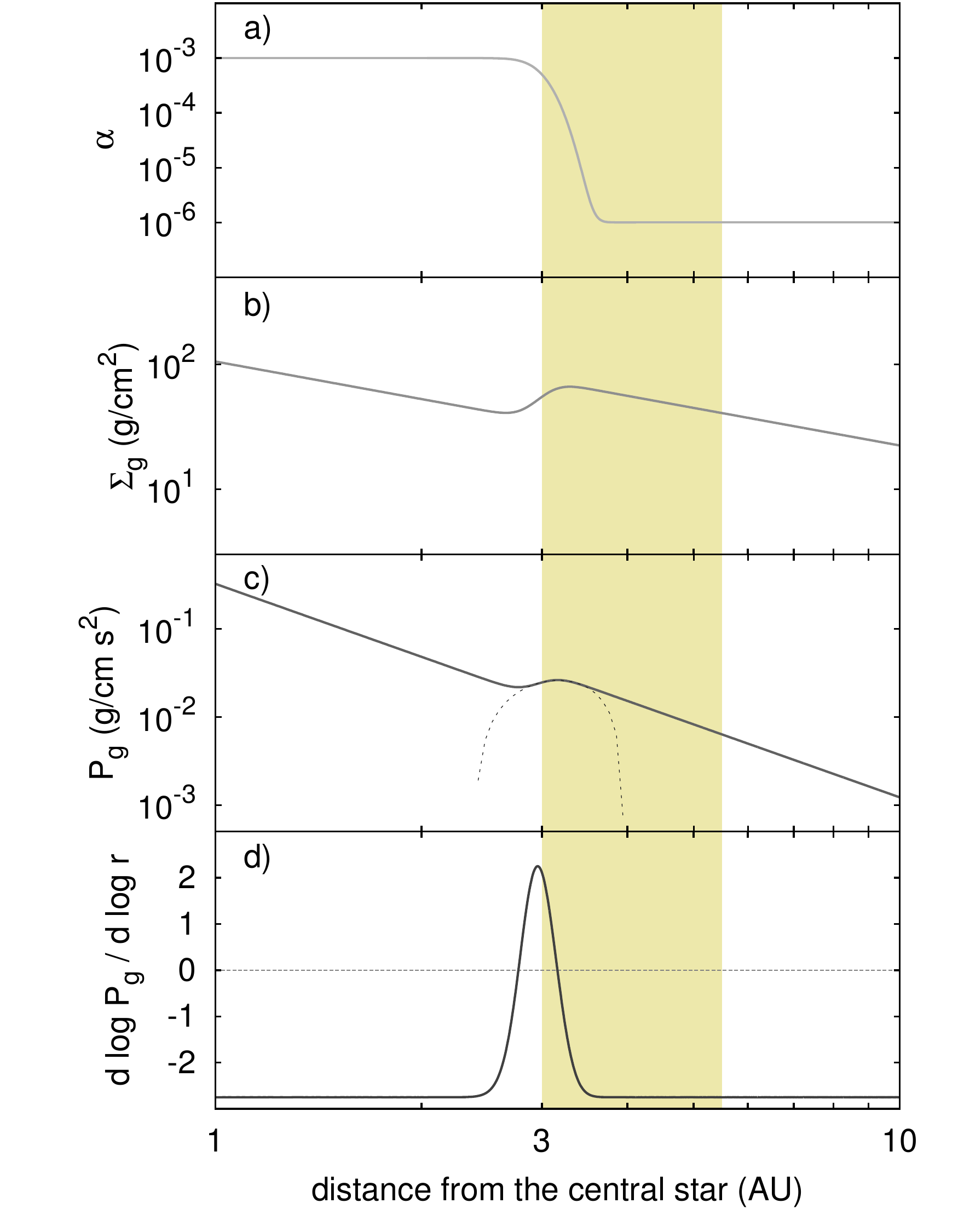}
      \caption{The disk model with the pressure bump near the snow line according to \citet{2007ApJ...664L..55K}. The panels show: a) the $\alpha$ parameter, b) gas surface density, c) gas pressure in the midplane and its Taylor expansion around the pressure bump location (Eq.\ (\ref{pg}), dashed line), d) gas pressure gradient, as a functions of the radial distance from the central star, for our fiducial disk model. Region highlighted with the different background color refers to models described in Sect.\ \ref{sub:2D}.
              }
         \label{fig:disk}
\end{figure}

In this section we present a simple analytical prediction of width of the annulus formed by the trapped particles of given Stokes number and compare it to results of test runs.
We use a disk model based on the work of \citet{2007ApJ...664L..55K}, where the $\alpha$ parameter varies with $r$ due to changes in the gas ionization. As the MRI turbulent strength depends on the degree of coupling to the magnetic field, a change in the gas ionization will affect $\alpha$. The gas ionization fraction depends on the total surface area of dust particles \citep{2009ApJ...698.1122O}, and is therefore most affected if there is a significant population of small particles. \citet{2007ApJ...664L..55K} assumed all particles to be $\mu$m-sized, meaning that the gas ionization rate is simply proportional to the dust to gas ratio. Beyond the snow line, the dust density steeply increases as the water vapor condenses into solid grains, causing a decrease in $\alpha$ that builds up a pressure bump on the disk accretion timescale. \citet{2007ApJ...664L..55K} present a disk model parametrized in the framework of the $\alpha$-prescription for a steady state obtained via the described mechanism.
Our implementation of the model is presented in Fig.\ \ref{fig:disk}. The $\alpha$ parameter drops down from $10^{-3}$ inside the snow line to $10^{-6}$ in the dead zone. This causes the bump in the surface density and the change of the sign of the pressure gradient. In the region where the pressure gradient is positive, the particles drift outwards and can thus be trapped in a so-called pressure trap. \citet{2010MNRAS.402.2436Y} remarked that in such model the local density maximum is Rayleigh unstable if the bump width is less than the disk scale height. Therefore, we choose the parameters of the model such that the width of the gas density bump measured by fitting a Gaussian is equal to 4 times gas pressure scale height.

The estimation of the trapped dust region width $L({\rm{St}})$ is done in a similar way as the derivation of the $h_{\rm{d,1}}({\rm{St}})$ in the previous section. We compare the timescales of the radial drift $\tau_{\rm{drift}}$ and turbulent diffusion $\tau_{\rm{diff}}$. As previously, we estimate the $\tau_{\rm{diff}}$ with Eq.\ (\ref{difftime}). We assume that to be trapped, the particle has to drift from its current location $r$ to the position of the pressure trap $r_0$. Thus, the radial drift timescale can be written as
\begin{equation}\label{tdrift}
   \tau_{\rm{drift}} = \left|\frac{r-r_0}{v_{\rm{drift}}}\right|,
\end{equation}
where the drift velocity $v_{\rm{drift}}$ can be obtained from Eqs (\ref{vdrift2})-(\ref{veta}), and it is proportional to the pressure gradient $\partial_{\rm{r}} P_{\rm{g}}$.
We assume that the disk is vertically isothermal, thus the gas pressure in the midplane is given by \citep{2007ApJ...664L..55K}
 \begin{equation}\label{pressure}
   P_{\rm{g}} = \frac{\Sigma_{\rm{g}}c_{\rm{s}}\Omega_{\rm{K}}}{2\pi}.
\end{equation}

In order to obtain the $L({\rm{St}})$, we want to get rid of the radial dependence of $\tau_{\rm{drift}}$. 
Thus, we approximate the pressure profile $P_{\rm{g}}(r)$ with the second order Taylor expansion around the location of the pressure bump $r_0$:
\begin{equation}\label{pg}
   P_{\rm{g}}(r) \approx P_{\rm{g}}(r_0) + \frac{1}{2} \frac{d^2{P_{\rm{g}}}}{dr^2}(r_0)\cdot(r-r_0)^2 = {\rm{C}}-{\rm{A}}\left(r-r_0\right)^2,
\end{equation}
and we find ${\rm{A}}\approx2\times10^{-28}$ g~cm$^{-\mathbf{3}}$~s$^{-2}$ and $\rm{C} \approx 2.6\times10^{-2}$~g~cm$^{-1}$~s$^{-2}$ for $r_0 \approx 3.16$~AU. The Taylor expansion is plotted with the dashed line in the panel c) of Fig.\ \ref{fig:disk}.
The derivative $\partial_{\rm{r}} P_{\rm{g}}$, needed to calculate the $v_{\rm{drift}}$, becomes
\begin{equation}\label{dpg}
   \partial_{\rm{r}} P_{\rm{g}} \propto -2{\rm{A}}(r-r_0).
\end{equation}
Thus, we can estimate the radial drift timescale $\tau_{\rm{drift}}$ as
\begin{equation}\label{tdrift2}
   \tau_{\rm{drift}} \approx \frac{\rho_{\rm{g}} \Omega_{\rm{K}}}{2{\rm{A}}} \left({\rm{St}} + \frac{1}{\rm{St}}\right).
\end{equation}

Comparing Eqs (\ref{difftime}) and (\ref{tdrift2}), using Eqs (\ref{ddust}) - (\ref{schmidt}), and replacing $\rho_{\rm{g}}c_s\slash\Omega_{\rm{K}} = \rho_{\rm{g}}H_{\rm{g}} = \Sigma_{\rm{g}}$ we find the expression for the width of the trapped dust annulus:
\begin{equation}\label{l}
   L \approx \left(\frac{\alpha}{\rm{A}}\Sigma_{\rm{g}}c_{\rm{s}}\Omega_{\rm{K}}\frac{1}{\rm{St}}\right)^{1\slash2}.
\end{equation}
One can notice that the width of the annulus becomes larger with growing surface density, turbulent viscosity or temperature of the gas, consistent with intuition.

\begin{figure}
   \centering
   \includegraphics[width=\hsize]{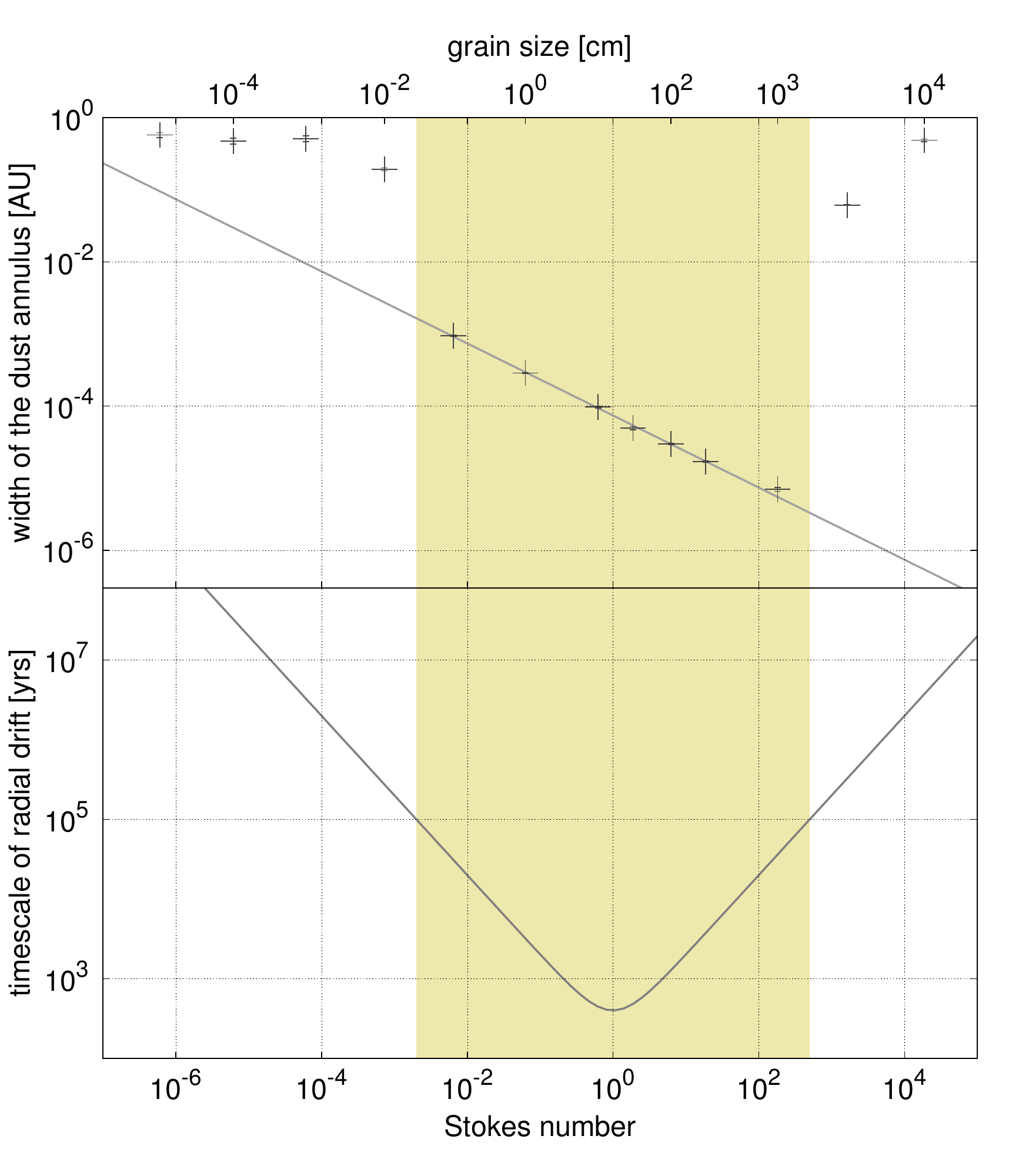}
      \caption{The top panel shows the analytically derived dependence for the trapped dust annulus width (Eq.\ (\ref{l}), line) and the results of test runs (points). The timescale of particles trapping is associated with the timescale of radial drift. The latter is specified on the lower plot (Eq.\ \ref{tdrift2}). The points on the top panel were measured after 10$^5$ years of evolution. This indicates a range of Stokes numbers of particles that can be trapped (marked with different background color).
              }
         \label{fig:trappingtest}
\end{figure}
The solids are trapped on a timescale of radial drift that is specified by Eq.\ (\ref{tdrift2}). The timescale is shortest for particles of ${\rm{St}} = 1$ and grows for both smaller and bigger sizes.
We perform a set of simulations using different sizes of particles ranging form 10$^{-5}$ to  10$^{4}$ cm. For each simulation we use 10$^5$ of representative particle distributed over 100 radial and 20 vertical zones. Initially the particles are placed between 3 and 4 AU. The collisions are switched off. After 10$^5$ yrs of evolution, the width of the bump in the dust surface density is measured by fitting Gaussian distribution. In the top panel of Fig.\ \ref{fig:trappingtest}, the fitted standard deviation of the distribution is plotted as a function of the Stokes number, together with the fit errors. In the bottom panel, the timescale of radial drift is shown. The range of Stokes numbers for that the timescale is shorter than $10^5$ yrs is indicated with different background color. The width of the trapped dust annulus on the top panel is consistent with the dependence given by Eq.\ (\ref{l}), but only for the particles in the range specified by the short enough timescale condition. This result is perfectly in agreement with our predictions.

In this test we neglect the radial drift velocity caused by gas accretion, specified by Eq.\ (\ref{vdrift1}). \citet{2012A&A...545A..81P} showed that if we do not neglect this effect, we get additional restriction for size of particles that can be trapped (their Eq.\ 11). Particles with Stokes number smaller than $\rm{St}_{crit}$ are not trapped because their coupling to gas is so strong that they move with the gas through the pressure maximum, where
\begin{equation}\label{stcrit}
  {\rm{St}_{crit}} = -\frac{v_{\rm{g}}^{\rm{r}}}{\partial_{\rm{r}} P_{\rm{g}}}\rho_{\rm{g}}\Omega_{\rm{K}},
\end{equation}
with $v_{\rm{g}}^{\rm{r}}$ as the radial velocity of gas. This condition holds only when the other component of the dust radial velocity is positive, i.e.\ $\partial_{\rm{r}} P_{\rm{g}} > 0$. In our model $\rm{St}_{crit} \approx 10^{-4}$, so this effect would not change the test result.

\section{Sedimentation driven coagulation}\label{sub:1D}
The Gaussian vertical structure of the dust as described in Sect.\ \ref{sub:vert} is usually a good approximation in the case of protoplanetary disk. However, it can be strongly affected by collisional evolution of the aggregates. 

We investigate the growth of the dust aggregates in a 1D vertical column. We base on a model presented by \citet{2005A&A...434..971D} and reproduced by \citet{2011A&A...534A..73Z} (henceforth ZsD11). The column is placed at the distance $r=1$ AU from the star of mass $M_\star = 0.5M_\odot$, with a gas surface density $\Sigma_{\rm{g}}=100$ g cm$^{-2}$ and a gas temperature $T_{\rm{g}}=200$~K. The radial drift is switched off. The dust particles are initially equal size monomers with radii $a_0=0.55$~$\mu$m and material density $\rho_{\rm{p}} = 1.6$ g cm$^{-3}$. They are initially vertically distributed such that the dust to gas ratio $\rho_{\rm{d}}\slash\rho_{\rm{g}}=0.01$ is constant along the column. Fragmentation is not included in this model, i.e.\ the particles collisions result in sticking for every collision energy. The growth is driven only by Brownian motion and differential settling. We ignore other sources of relative velocity: radial and azimuthal drift as well as turbulence. For the test we used $5\times10^4$ representative particles and 100 cells (500 particles per cell). The test run took about 48 hours on an 8 core 3.1 GHz AMD processor.

Similar to \citet{2005A&A...434..971D} and ZsD11, we notice that initially the growth is slow, driven by the Brownian motions, and proceeds faster closer to the midplane, where the matter density is highest. At $t\approx100$ yrs, the particle growth in the upper layers speeds up as the differential settling comes into play. The value of the vertical settling velocity increases with height, as can be noticed from Eq.\ (\ref{vsvelmin}). The aggregates grow and settle down simultaneously. The first rain-out particles that reach the midplane have masses of around 10$^{-2}$ g. 

\begin{figure}
   \centering
   \includegraphics[width=\hsize]{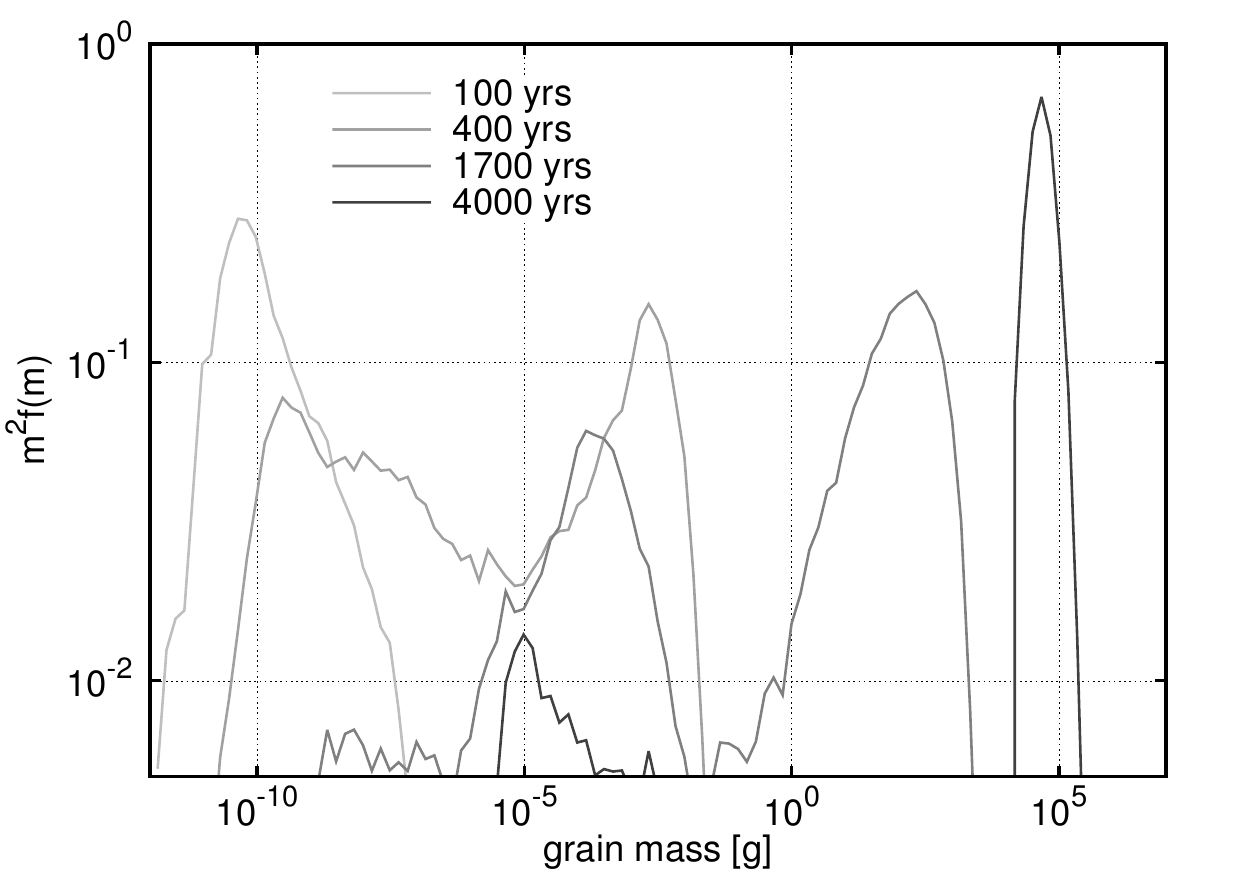}
      \caption{Vertically integrated dust mass distribution at different stages of evolution for the sedimentation driven coagulation test. After approximately 400 yrs, the dust distribution splits into two parts. The big aggregates continue to grow at the expense of the small particles.
              }
         \label{fig:aghist}
\end{figure}
\begin{figure}
   \centering
   \includegraphics[width=0.93\hsize]{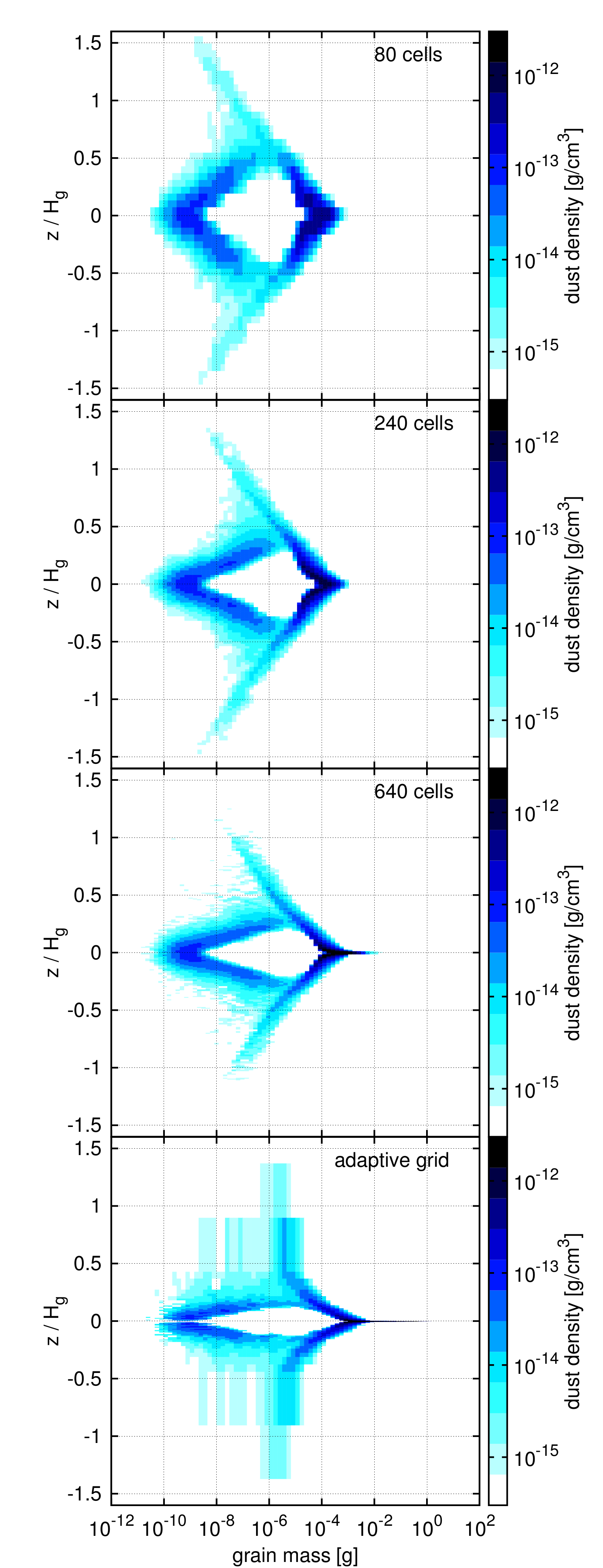}
      \caption{The vertical distribution of the dust grains of different sizes for the 1D sedimentation driven coagulation test. The three upper panels show the result of simulations with constant grid with growing number of cells. The bottom panel uses the adaptive grid routine described in this work with 100 cells. All the distributions were plotted after 1000 yrs of the evolution. The numerical convergence of results is noticeable.
              }
         \label{fig:dd05}
\end{figure}
Fig.\ \ref{fig:aghist} presents the mass distribution evolution obtained in this test. 
It can be noticed that within the first 400 years the dust distribution becomes bimodal. One population consists of the rain-out particles, which reached the midplane, and the other one are the smaller aggregates, which remain vertically dispersed. The bigger particles grow at the expense of the small ones, thus the surface density of the latter decreases.
The final mass of the biggest aggregates is $\sim$10$^6$ g. 
Such a bimodal dust distribution for sedimentation driven coagulation was also reported by \citet{2005A&A...434..971D} and \citet{2005ApJ...625..414T}.

The numerical model used by ZsD11 is practically identical to ours, but the spatial grid is fixed and consists of equally spaced cells, while in our case we use the adaptive grid method. They use 40 cells to resolve 4 gas pressure scale heights. We notice that in comparison to their results, we get a faster evolution of the dust. The first rain-out particles arrive to the midplane after approximately 400~yrs of evolution, instead of 500~yrs reported by ZsD11. We observe also that the growth proceeds to bigger sizes than in ZsD11, where the growth stalls at approximately 10$^{-1}$~g.

In order to explain the discrepancy of the results obtained by ZsD11 and us, we perform resolution test. 
As the adaptive grid reflects a very high number of cells in high density regions, we investigate if the result obtained by ZsD11 depends on the number of cells used. Therefore, we perform a set of simulation with constant, equally spaced grid but increasing the number of cells. Fig.\ \ref{fig:dd05} presents the mass-height distribution of the dust after 1000 yrs of evolution for the constant grid with 80, 240 and 640 cells as well as for the adaptive gridding with 100 cells. Note that ZsD11 modeled only the upper half of the column, so their 40 cells is equivalent to our 80 cells resolution.
 We find that the timescale of the evolution is indeed dependent on the vertical resolution. With the adaptive grid method, we are also able to see the effect of sweeping up of the small particles by the big ones on the dust distribution around the midplane (see the bottom panel of Fig.~\ref{fig:dd05}).
 
If we consider one grid cell with a bottom wall at $z=0$, using the model described in this section, the collision rate defined by Eqs (\ref{rik})-(\ref{Kik}) does not depend directly on the height above the midplane of the center of the cell $z_{\rm{c}}$. The relative velocity $\Delta v$ is dominated by the differential settling velocity that is directly proportional to $z_{\rm{c}}$. Also the cell volume $V$ is directly proportional to $z_{\rm{c}}$. Therefore, one could expect that the collisional evolution does not depend on the vertical resolution we choose. However, we find that the higher resolution we use, the faster the growth and settling proceed. This effect can be explained in the following way: we calculate the relative velocities of particles basing on the physical values obtained in the centers of the cells. Thus, the exact values of gas density, Stokes numbers and vertical settling velocities depend on the exact choice of the location of the cell. All these values influence the collision rate of particles. The more cells we use, the closer to the midplane (where the growth proceeds fastest at the very beginning as well as at the end of the evolution) we are able to resolve. On the other hand, the faster growth we obtain, the quicker the particles settle down.

It is worth noting, that one of the basic assumptions of the method we use \citep{2008A&A...489..931Z} is that the particles are homogeneously distributed over the volume of the cell within they can collide. If we do not assure sufficiently high spatial resolution, this assumption is broken, and the model leads to unphysical results.

The vertical resolution defines the maximum dust to gas ratio we are able to obtain. In the case of constant grid with the number $N_{\rm{c}}$ of cells the maximum dust to gas ratio $\rho_{\rm{d}}\slash\rho_{\rm{g}}=N_{\rm{c}}\times0.01$ would occur if we place all of the dust particles in one cell. The 0.01 is the global dust to gas mass ratio. In the case of the adaptive grid the dependence on the number of cells is much weaker, and we are able to resolve higher dust to gas ratios with much lower number of cells. 

The impact of the dust layer width on the growth was investigated by \citet{1986Icar...67..375N}. They concluded that the growth terminates for an infinitely thin layer, as when all of the bodies are located at $z=0$ the vertical velocity of dust resulting from Eq.\ (\ref{vsvelmin}) $v_{\rm{d}}^{z}=0$, and the main source of the relative velocities driving the collisions vanishes. However, even with the adaptive grid, we can never obtain an infinitely small cell, so the growth termination does not occur in our simulations. 

The existence of such an infinitely thin dust layer is unrealistic anyway. As soon as the dust to gas ratio exceeds unity, the shear instabilities \citep{1980Icar...44..172W,1993Icar..106..102C}, in particular the Kelvin-Helmholtz instability \citep{2006ApJ...643.1219J} are known to occur. \citet{2010ApJ...722.1437B} showed that in the case of no turbulence, another kind of hydrodynamic instability, namely the streaming instability \citep{2005ApJ...620..459Y}, will generate a turbulence and maintain the dust to gas ratio before the Kelvin-Helmholtz instability could be triggered. It then prevents the dust from further settling and the growth from terminating.

With the adaptive grid routine, we are able to resolve the dust to gas ratio much higher than one. 
To avoid such an unphysical situation, we implement an artificial $\alpha$ viscosity, $\alpha_{\rm{SI}}$, that is designed to mimic the impact of the streaming instability on the vertical distribution of dust. We calculate the $\alpha_{\rm{SI}}$ as
\begin{equation}\label{alphaSI}
   \alpha_{\rm{SI}} = \alpha_{\rm{SI}, max} \left[ 1 + \erf{ \left(\frac{\rho_{\rm{d}}\slash\rho_{\rm{g}} - {\rm{c_1}}}{\rm{c_2}}\right) } \right],
\end{equation}
where $\alpha_{\rm{SI}, max}$ defines a minimal turbulent viscosity that we need to maintain dust to gas ratio lower than one and the $\rm{c_1}$, $\rm{c_2}$ are parameters of the error function $\erf$.
The $\alpha_{\rm{SI}, max}$ can be calculated from Eq.\ (\ref{hdust}) as 
\begin{equation}
   \alpha_{\rm{SI}, max} =  Z_0^2 \min(0.5,{\bar{\rm{St}}})\left(1+\bar{\rm{St}}^2\right),
\end{equation}
with $Z_0$ representing initial dust to gas ratio, and $\rm{\bar{St}}$ being the Stokes number averaged over all particles present in given cell, as the strength of the streaming instability driven turbulence is determined by the collective property of the particles.
The form of Eq.\ (\ref{alphaSI}) was chosen such that the additional term of viscosity is nonzero only when the dust to gas ratio exceeds unity and it adds only the amount of turbulence that is needed to maintain the dust to gas ratio below the unphysical value.

\begin{figure}
   \centering
   \includegraphics[width=1.1\hsize]{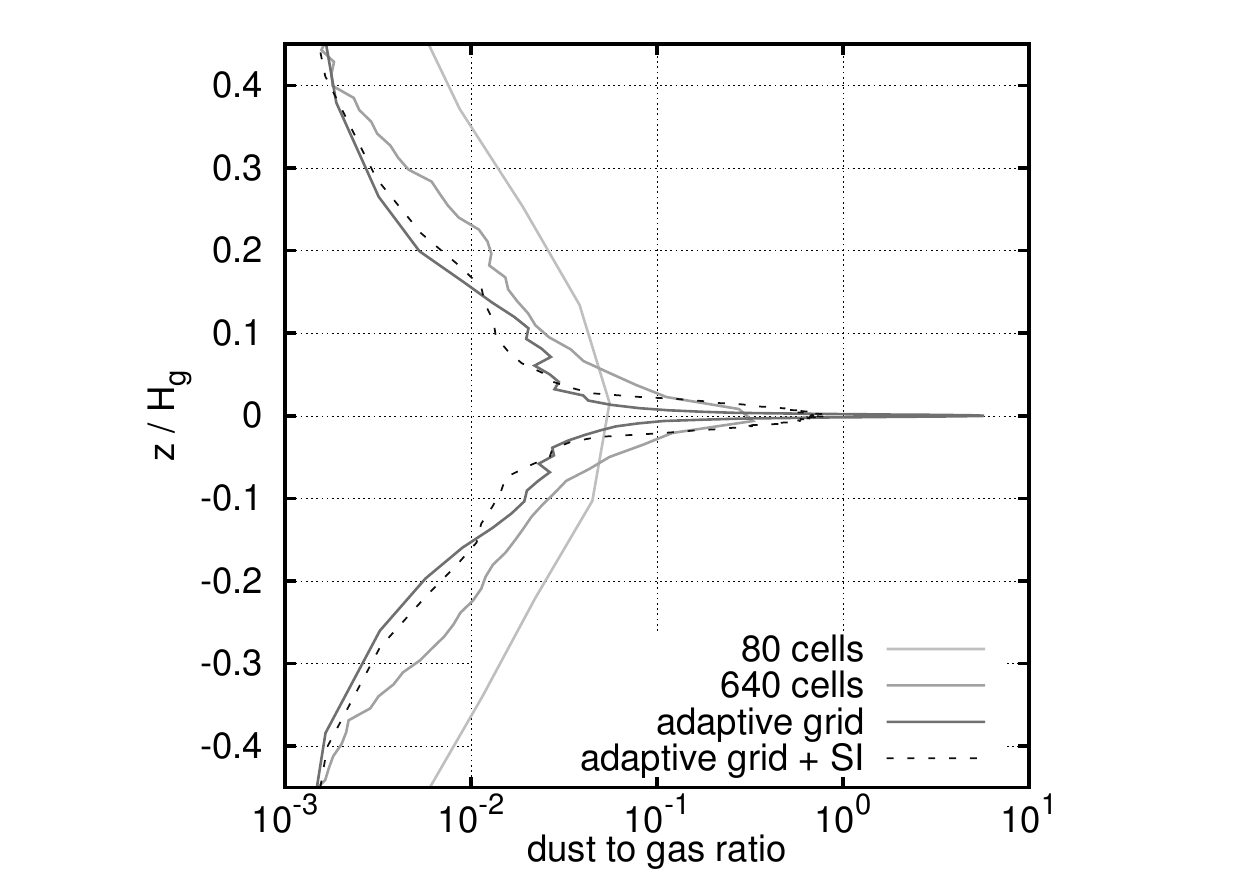}
      \caption{The dust to gas ratio around the midplane after 1000 yrs of evolution as resolved by different algorithms: constant grid with 80 and 640 cells and the adaptive grid with 100 cells with and without the streaming instability (SI). The obtained dust to gas ratio depends strongly on the vertical resolution. In the case of the adaptive grid it exceeds unity. The implementation of the streaming instability lowers the dust to gas ratio only in the region in that such an unphysical values occur.
              }
         \label{fig:dusttogas}
\end{figure}
The Fig.\ \ref{fig:dusttogas} shows the dust to gas ratio at different height above the midplane after 1000 yrs of evolution for the different resolutions and for the test with the artificial viscosity introduced by the $\alpha_{\rm{SI}}$. The resolution dependence can be noticed. The dust to gas ratio in the case of the adaptive grid exceeds unity. The implementation of the $\alpha_{\rm{SI}}$ changes the dust to gas ratio only very close to the midplane.

We find that implementing such an additional turbulence source speeds up the coagulation of the big particles. This is because it increases the relative velocities of the bodies and thus the collision rates. As we ignore the possibility of the aggregates fragmentation, the higher relative velocities result in faster growth.

\begin{figure}
   \centering
   \includegraphics[width=\hsize]{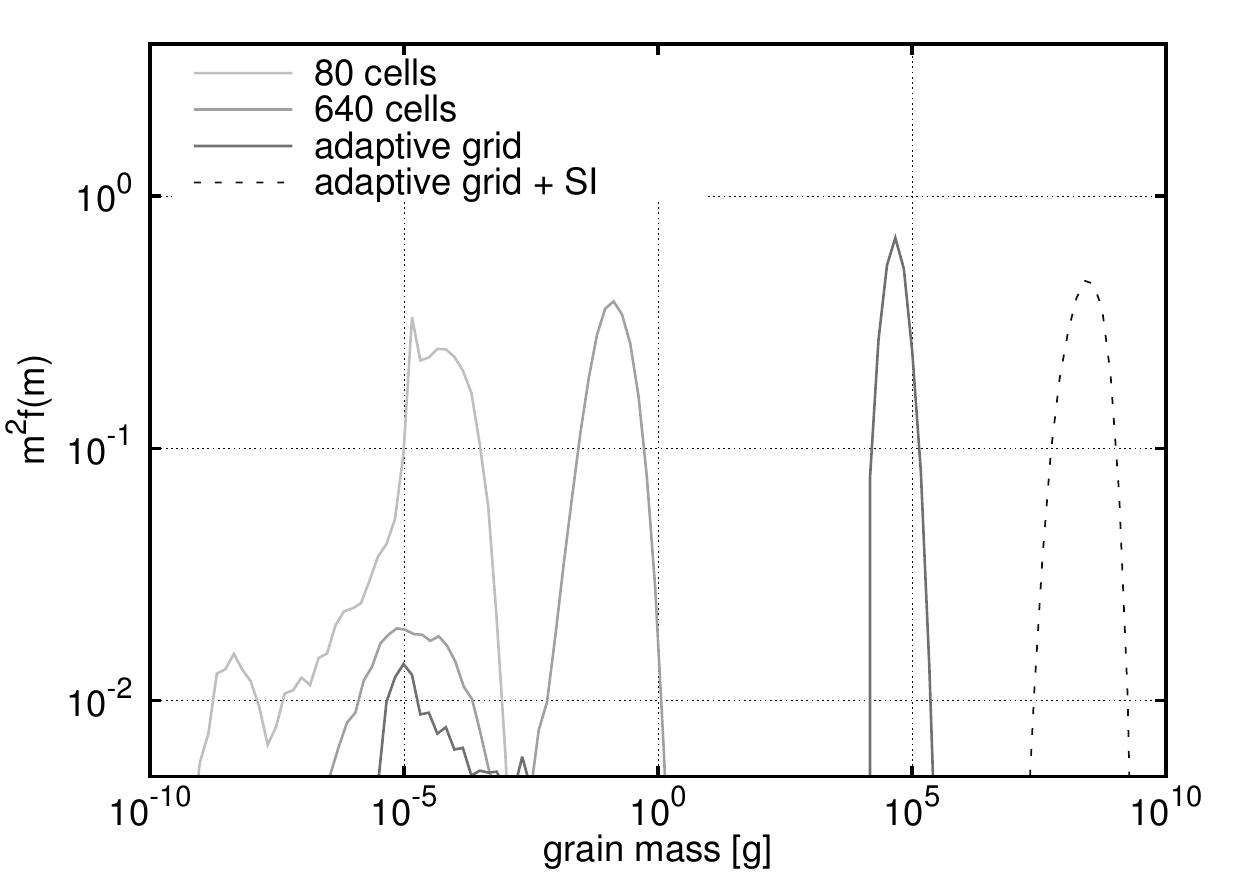}
      \caption{The dust mass distribution after 4000 yrs of evolution for the tests with the constant grid with 80 and 640 cells and adaptive gridding with 100 cells with and without the streaming instability (SI) implemented. The figure reveal a huge impact of the vertical resolution on the dust growth timescale. With the adaptive gridding we obtain much bigger bodies after the same time of evolution. Taking the SI into account additionally speeds up the growth.
              }
         \label{fig:histcomp}
\end{figure}
The Fig.\ \ref{fig:histcomp} shows the mass distributions after 4000 yrs of evolution obtained for different gridding as well as with and without the streaming instability (SI). This figure reveals how much the growth depends on the resolution. The mass of the biggest agglomerates obtained after 4000 yrs in the test with 640 constant cells and 100 adaptive grid cells varies by five orders of magnitude. This is however a timescale effect. If we wait long enough, which is of the order of Myrs for the constant grid, we will obtain the same resulting size of agglomerates. The growth can proceed only until all the small particles are swept up by the big ones.

The consideration of the streaming instability allows us to obtain another 4 orders of magnitude in mass larger particles. The additional viscosity increases the vertical extent of the big bodies as well as their relative velocities. Thus, they are able to collide with the small particles that reside higher above the midplane. This speed up of the growth can however be a result of the simplified instability implementation we used. We do not account for the strong particle clumping reported for the streaming instability \citep{2009ApJ...704L..75J}. We ignore also the possibility of the gravitational instability of the clumps \citep{2007Natur.448.1022J}. We plan to include these effects in our future work.

The growth timescale dependence on the vertical resolution revealed in these sections can have a huge impact on dust evolution models. In 2D cases the impact of vertical structure resolution is even stronger as the relative velocity is dominated by the radial and azimuthal drift. Its value does not depend on the vertical position, so the collision rate becomes explicitly dependent on $z$. 

In the model presented here, turbulence is not included, besides the one generated by the streaming instability. We ignore also the possibility of the aggregates fragmentation. We expect that including these effects would lower the discrepancy between the results obtained using the constant and adaptive gridding. The turbulent mixing prevents the high dust to gas ratio, which is problematic for the constant grid scheme. Taking the fragmentation into account sets up a maximum mass over which the particles cannot grow. Thus, the difference in the growth timescale does not change the maximum size of particles after the same time of evolution obtained in the different models.

The tests presented in this section have proven our adaptive grid to deal very well with the high dust density regions. However, as can be observed on the bottom panel of Fig.\ \ref{fig:dd05}, the low density regions are resolved much worse. As the most of the coagulation happens in the high density regions around the midplane, this flaw should not effect the mass distribution function evolution. However, it may limit the possibilities of using our code in context of the protoplanetary disks observations.

\section{Sweep-up growth at the inner edge of dead zone}
\label{sub:2D}

\begin{figure}
   \centering
   \includegraphics[width=\hsize]{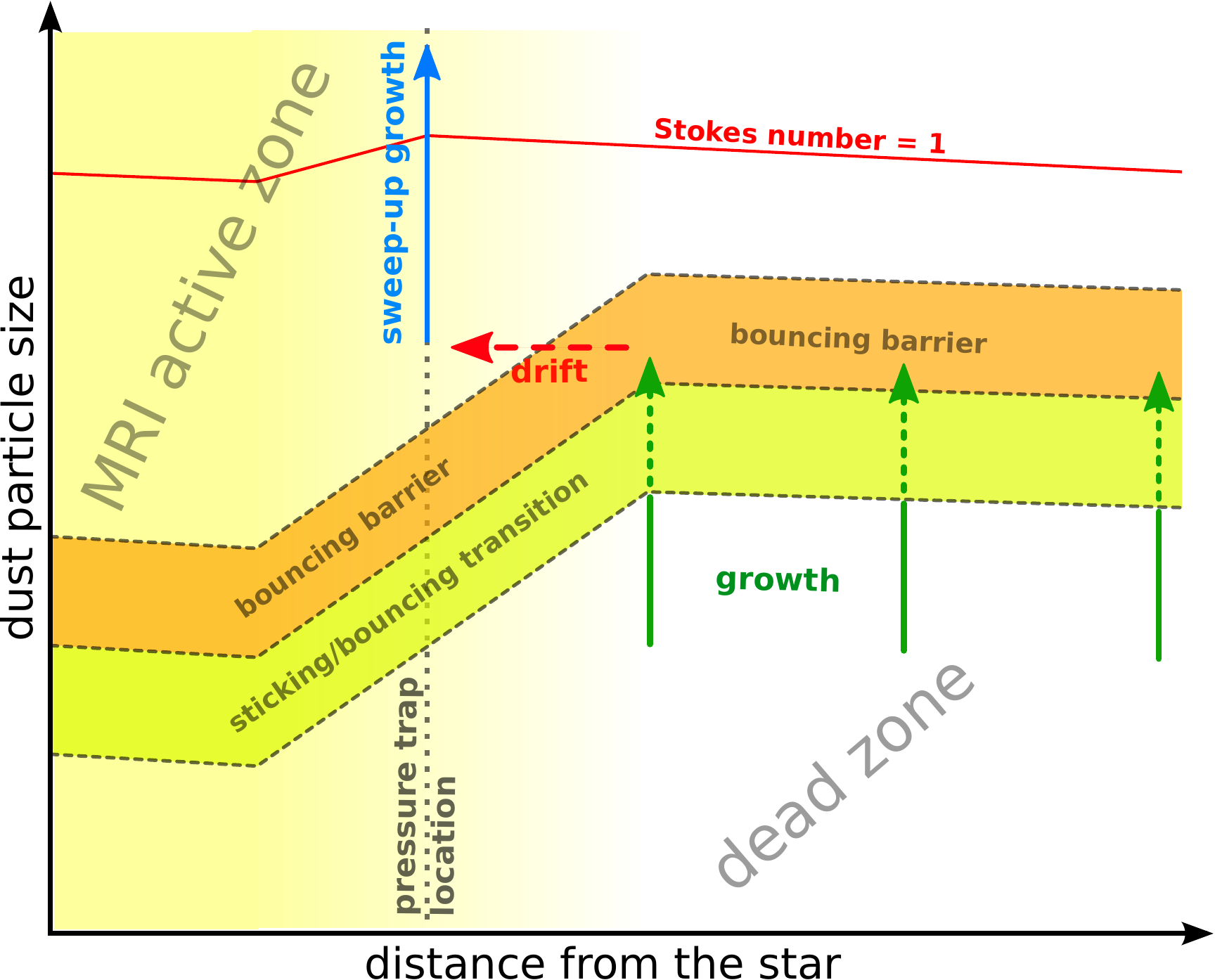}
      \caption{Sketch of the planetesimal formation mechanism we suggest. 
      Thanks to the radial variation in turbulence efficiency, the position of the collision regimes is shifted in terms of particle size. 
      The dust aggregates can grow to larger sizes in the dead zone than in the MRI active zone.
      The “big” particles grown in the dead zone drift inwards through the bouncing regime, to the location of the pressure trap, and some of them can continue to grow via sweeping up the small particles halted by the bouncing barrier.
              }
         \label{fig:sketch}
\end{figure}

In the test models presented so far, we have always assumed perfect sticking between particles, and ignored all other possible collision outcomes. However, collisional physics of dust aggregates is highly complex, and laboratory experiments have shown that also effects such as bouncing, fragmentation and erosion can occur \citep{2008ARA&A..46...21B}. By implementing the collision scheme proposed by \citet{2010A&A...513A..56G} in 0D simulations, \citet{2010A&A...513A..57Z} showed the importance of using a realistic collision model, and discovered the existence of the bouncing barrier, where growth-neutral bouncing collisions can completely prevent particle growth above millimeter-sizes, even before the fragmentation barrier is reached.

\citet{2012A&A...540A..73W} showed that the existence of a collisional growth barrier (such as the bouncing barrier) can actually be beneficial for the growth of planetesimals. If some larger particles, or {\it seeds}, are artificially introduced into a 0D model, they can grow by sweeping up the population of particles kept small by the bouncing barrier, thanks to the mass transfer effect observed by \citet{2005Icar..178..253W}. When two large particles collide at a high velocity, they tend to fragment, but if the mass ratio between the colliding particles is high enough, only the smaller of the two will be disrupted, depositing a fraction of its mass onto the larger particle in the process. In this way, two populations of particles are formed, where the few seeds grown by sweeping up the small particles while colliding only rarely between themselves. \citet{2012A&A...544L..16W} and \citet{2013ApJ...764..146G} showed that the first seeds might be formed by including velocity distributions produced by stochastic turbulence, but the exact nature of these distributions, and whether the effect is capable of producing high enough mass ratios, is still unclear.

In this study, we investigate whether the seeds can be produced at one location in the disk and then transported by the radial drift to another region, where they are significantly larger than the grains produced locally. This could allow them to grow further by sweeping up the smaller grains. In particular, we postulate that such a situation can occur for a sharp $\alpha$ change, e.g. at the inner edge of a dead zone.

Fig.~\ref{fig:sketch} shows the basic idea behind our model. 
At the inner edge of the dead zone, strength of the MRI turbulence drops, affecting the relative velocities between dust particles. In the MRI active region, the turbulence is stronger than in the dead zone, causing bouncing to occur for significantly smaller particles. Thus, aggregates growing in the dead zone can reach larger sizes. The radial drift (that increases toward the Stokes number equal unity) can transport the largest particles to the MRI active region, and at the same time into another collisional regime. The drifting particles have now become seeds, and can continue to grow by sweeping up the small grains stuck below the bouncing barrier. Furthermore, the rapid turbulence strength decline can result in a formation of a pressure trap that allows the seeds to avoid further inward drift and becoming lost inside of the evaporation radius of the star.

A difficulty in the planetesimal formation by sweep-up growth scenario is that the first seed particles have to be orders of magnitude more massive than the main population, and that if too many such seeds are formed, they will fragment among themselves too often to be able to grow. As a first application of our 2D code, we investigate whether the planetesimal formation via the mechanism described above can be initiated in a realistic protoplanetary disk.

We focus on a protoplanetary disk with a pressure bump around the snow line \citep{2007ApJ...664L..55K}, using the disk model presented in Sect.~\ref{sub:trap}. We assume a stationary disk, which is a simplification, as the dust grains size distribution, evolution of which we model, should affect the disk structure. We discuss this issue further in Sect.\ \ref{sub:last}. The total disk mass integrated between 0.1 and 100 AU is $0.01M_{\odot}$, and we set the mass accretion rate to 10$^{-9}$ $M_{\odot}$~yr$^{-1}$. Thus, our model corresponds to a low-mass, passive protoplanetary disk. We focus this study on the region around the pressure bump, between $r = 3-5.5$ AU, as highlighted in Fig.~\ref{fig:disk}. At 3~AU, the disk has a gas surface density $\Sigma_{\rm g} = 65$ g cm$^{-3}$ and a temperature $T_{\rm{g}} = 140$~K. This disk model is highly simplified, especially in the outer regions, but as we focus only on the inner region, we consider it a good approximation. We also assume a stationary gas disk, since, because of the computational expense of the simulations, we only run the models for $\sim$$3 \times 10^4$ yrs, which is much shorter than the typical disk evolution timescale.

We assume an initial dust to gas ratio of $0.01$, and distribute the dust mass into monomers of size $a_0=1$~$\mu$m. The internal density of the particles is set to $\rho_{\rm{p}}=$ 1.6~g~cm$^{-3}$. For the models presented in this study, we use over a half a million (exactly 2$^{19}$) representative particles and an adaptive grid resolution of 64 radial and 32 vertical zones. This gives 256 representative particles per cell, which allows us to resolve the coagulation physics properly (see Sect.\ \ref{sub:test1}). Each swarm represents $\sim$$10^{22}$~g, corresponding to a maximum representative particle size of roughly 100 km that is obtainable without breaking the requirement that the number of representative particles must be lower than the number of physical particles in the swarm they represent. At the current stage of our project we do not reach km-sizes.

\begin{figure}
   \centering
   \includegraphics[width=\hsize]{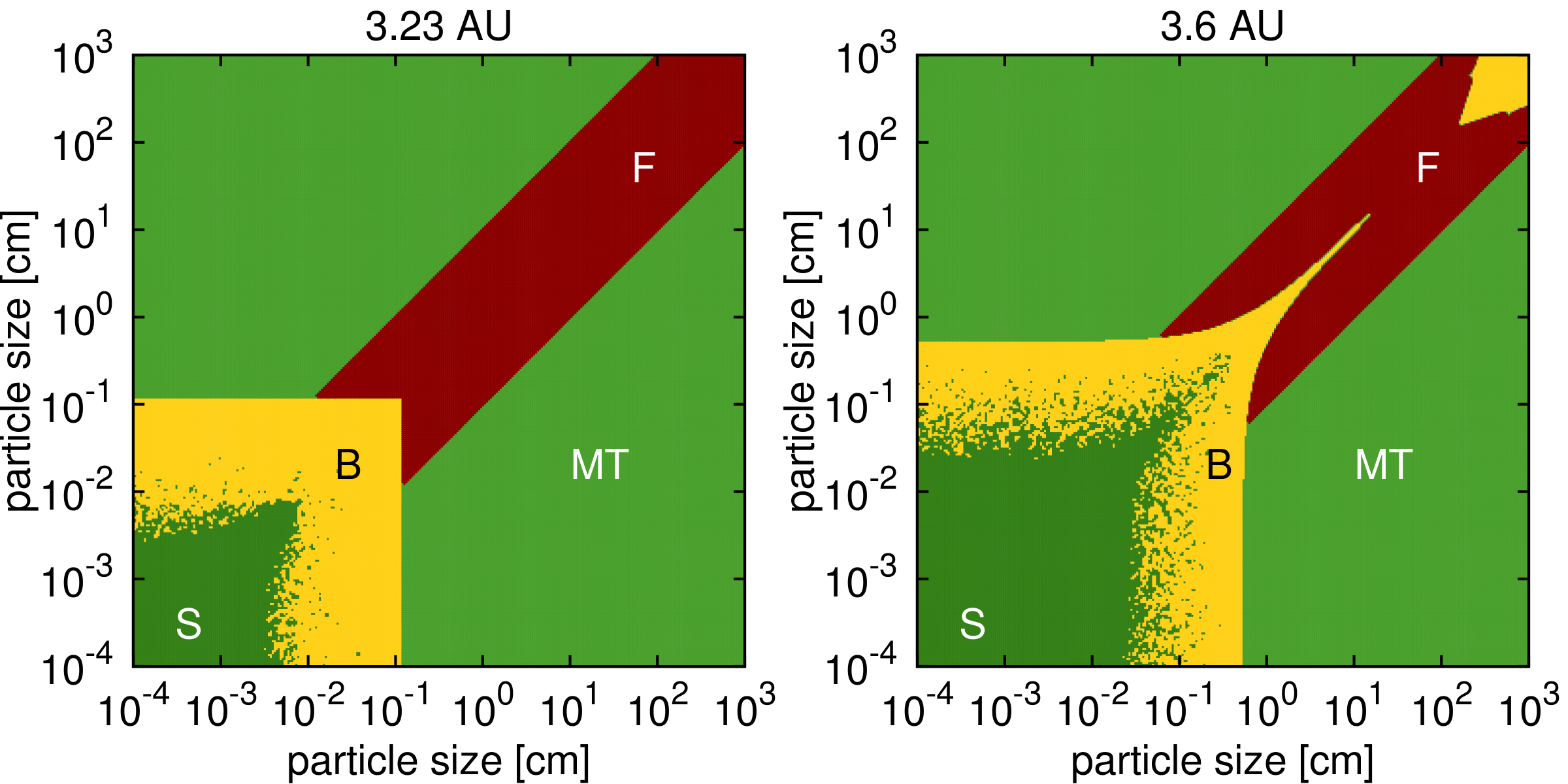}
      \caption{The collision outcome for particle pairs of given sizes located in the midplane at 3.23 AU (pressure trap) and at 3.6 AU (dead zone) in collision model B. “S” marks sticking, “B” bouncing, “F” fragmentation, and “MT” mass transfer. Thanks to the change in the disk properties between the two regions, the bouncing barrier occur at different particle sizes, as predicted when constructing our planetesimal formation scenario (Fig. \ref{fig:sketch}).}
         \label{fig:colscheme}
\end{figure}
\begin{figure*}
   \centering
   \includegraphics[width=\hsize]{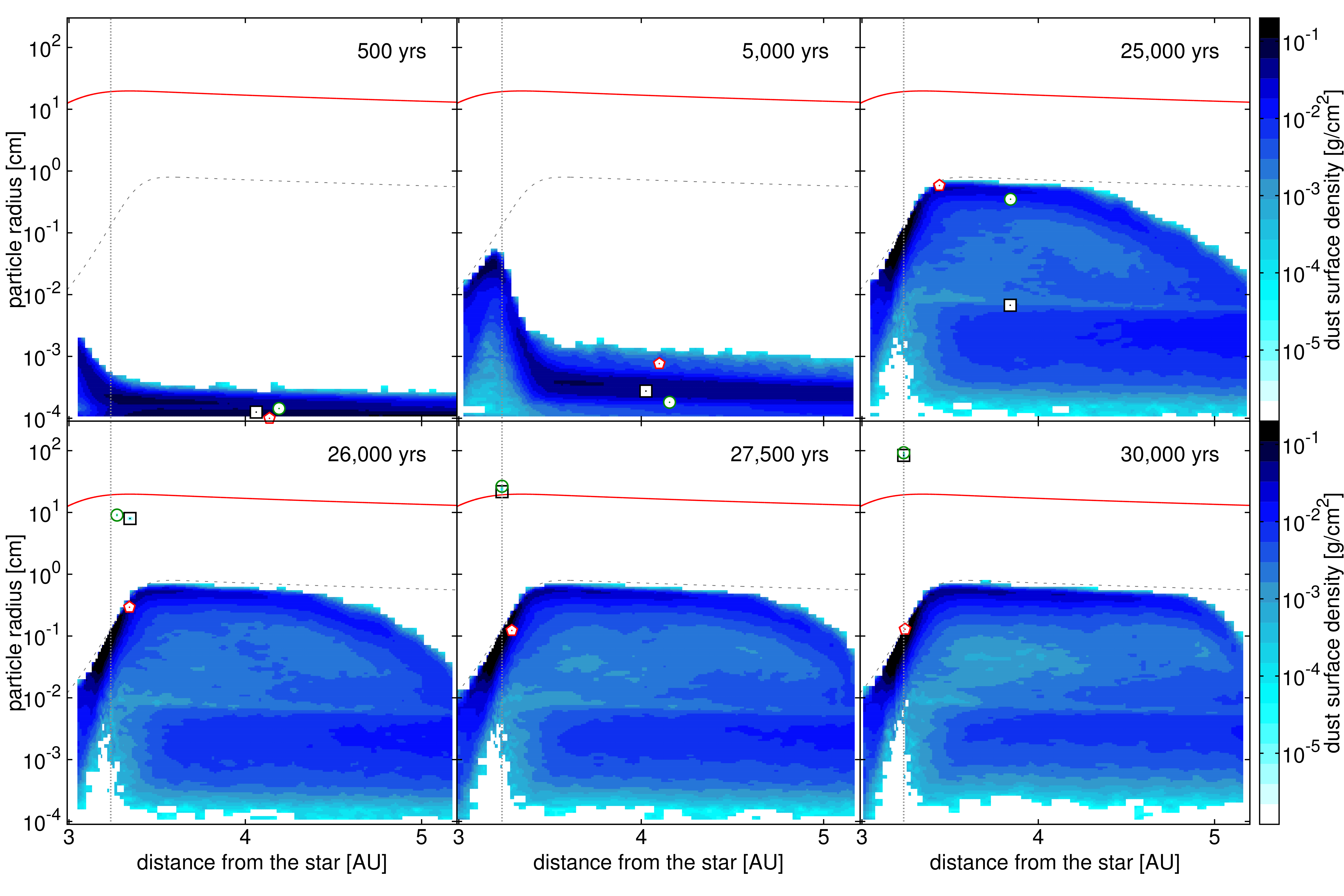}
   \caption{The vertically integrated dust density at different stages of the evolution, using collision model B. The solid line shows the particle size corresponding to the Stokes number of unity, where the drift is the fastest. This line is also proportional to the gas surface density. The dashed line shows approximate position of the bouncing barrier. The dotted line indicates the location of the pressure trap. The symbols point the position of three selected representative particles. Two of them are the particles that become the seeds that continue growing, trapped in the pressure trap at 3.23~AU, while the growth of the other swarms is stopped by the bouncing barrier. The feature at $r>3.5$~AU and $a<$10$^{-2}$~cm comes from the bimodal distribution revealed in the 1D tests (see Sect.\ \ref{sub:1D}), where the small particles are vertically dispersed, while the bigger particles resides in the midplane of the disk.}
          \label{fig:densar_mt}%
\end{figure*}
For our collision model, we use two simplified prescriptions (models A and B) for bouncing, fragmentation and mass transfer based on the work of \citet{2012A&A...540A..73W}. In both models, we determine the sticking probability as a function of the relative velocity $\Delta v$:
\begin{equation}
p_{\rm s}(\Delta v) = \left\{ \begin{array}{ccl}
1 		&		& \Delta v < v_{\rm s} \\
0 		& {\rm if} 	& \Delta v > v_{\rm b} \\
1-k 		&		& {\rm otherwise}, \\
       \end{array} \right.
\end{equation}
where $k = \log(1 + \Delta v - v_{\rm s}) /\log(1 + v_{\rm b} - v_{\rm s})$, consistent with the findings by \citet{2012Icar..218..688W}. The smooth transition between sticking and bouncing collisions turns out to be a natural way to limit the number of potential seeds, i.e. particles that are large enough to initiate sweep-up in the dead zone.

The fragmentation probability is determined by a step function:
\begin{equation}
p_{\rm f}(\Delta v) = \left\{ \begin{array}{ccl}
0 		& {\rm if}	& \Delta v < v_{\rm f} \\
1 		& 	 	& \Delta v \geq v_{\rm f}, \\
       \end{array} \right.
\end{equation}
and we let $v_{\rm s} = 3$~cm~s$^{-1}$, $v_{\rm b} = 60$~cm~s$^{-1}$, and $v_{\rm f} = 80$~cm~s$^{-1}$ be the sticking, bouncing and fragmentation threshold velocities. These values correspond to silicate grains, which are believed to be less sticky and resilient to fragmentation than icy grains that would also exist in the simulation domain. However, because of the lack of knowledge about the ice collision properties, and the uncertainty in the efficiency in sublimation and sintering at the snow line, we decide to take the pessimistic approach of using only the silicates. The $v_{\rm s}$, $v_{\rm b}$ and $v_{\rm f}$ are here independent on particle masses and $\Delta v$, which is different from the \citet{2012A&A...540A..73W} model. This is a significant simplification coming from the code optimization reason. However, the order of magnitude of these values is consistent with the original model, thus the overall scheme of collisional evolution is preserved.

In both of the models, during a fragmenting event, the mass of both particles is distributed according the power-law $n(m) \propto m^{-9/8}$, consistent with findings by \citet{1993Icar..106..151B} as well as \citet{2010A&A...513A..56G}, and the representative particle is selected randomly from the fragments (see \citet{2010A&A...513A..57Z} for details on how this is done in the representative particles and Monte Carlo fashion).

In model B, we also include the mass transfer effect, which occurs during a fragmenting event when the particle mass ratio is high enough, namely $m_1/m_2 > m_{\rm crit}$ ($m_1>m_2$), where we put $m_{\rm crit} = 10^3$. We assume a constant mass transfer efficiency of $0.8\cdot m_2$, i.e.\ the more massive particle gains 80\% of the mass of the smaller particle. 

The collisional model developed by \citet{2012A&A...540A..73W} is much more complex than ours. In their work the mass transfer efficiency is dependent on the impact velocity. We decided to assume the mass transfer efficiency to be constant, as we are here mostly concerned at the point where sweep-up is initiated, and we do not want to model the process in detail. For the same reason we ignored the threshold between erosion and mass transfer that \citet{2012A&A...540A..73W} found to be important for the growth to planetesimal sizes.

In Fig.~\ref{fig:colscheme}, we present the collision outcome for all particle pairs with collision model B, in the midplane, at both the location of the pressure trap (3.23~AU) and in the dead zone (3.6~AU). In the case of collision model A, the plot is similar, but the mass transfer regime is replaced by fragmentation.
From the plot, we can notice that due to differences in turbulent viscosity, the bouncing and fragmentation occur at different sizes depending on the location of the disk. In the dead zone, the turbulence is extremely low, $\alpha=10^{-6}$, compared to $\alpha \approx 10^{-4}$ at the pressure trap, and the particle growth can therefore continue to more than one order of magnitude larger sizes before the bouncing barrier halts it. This is exactly what is needed for our planetesimal formation via sweep-up mechanism to work.

In the case of collision model A, the growth is halted by the bouncing barrier at $\sim$0.1 cm in the pressure trap region and $\sim$0.7 cm in the dead zone, and there is no possibility that the growth could proceed towards bigger sizes. In model B, if radial drift would be ignored and the growth would only be allowed to proceed locally, the particle growth would stop at the same sizes as in model A. However, we find in our simulations that when both drift and mass transfer are included, the situation changes significantly, in a way that enables sweep-up, as discussed earlier.

The result of the simulation using collision model B is illustrated in Fig.~\ref{fig:densar_mt}, where we plot the vertically integrated dust density evolution at six different times between $t = 500$~yrs and $t = 30,000$~yrs. The dust growth proceeds the fastest in the inner part of the domain, where the relative velocities are the highest because of stronger turbulence. After $1,000$~yrs, the particles in the inner part of the disk have reached the bouncing barrier, which efficiently halts any further growth. The position of the bouncing barrier, indicated with the dashed line in Fig.~\ref{fig:densar_mt}, is estimated analogically as the location of the fragmentation barrier in \citet{2011A&A...525A..11B}. As time progresses, particles further out also halt their growth due to bouncing. The bouncing barrier occurs at larger sizes in the dead zone than in the pressure trap. After $20,000$ years, most of the particles are kept small by the bouncing, and only evolve by slowly drifting inwards. The size of the particles stopped by the bouncing in the dead zone corresponds to ${\rm{St}} < 5\times10^{-2}$, for which the drift timescale $> 10^4$ yrs (see Fig.\ \ref{fig:trappingtest}). Thus, the small dust is still present beyond the pressure trap at the end of the model.

During the inward drift, the particles halted by the bouncing barrier in the dead zone are automatically shifted to the fragmentation$\slash$mass transfer regime in the region of higher turbulence. 
Most of these particles fragment due to equal-size collisions, which can be seen in the Fig.~\ref{fig:densar_mt}, as the majority of the bigger particles from the dead zone is fragmented down to the position of the bouncing barrier.
With these contour plots, however it is not easy to display the minute, but very important, amount of bodies that are able to cross the barrier unscathed: these are the seeds. The Monte Carlo method finds 2 such seed representative particles in the model B run, which is hard to show in the contour plots, so that we mark them separately in Fig.~\ref{fig:densar_mt}. In the figure, we plot the exact positions of three selected swarms. All of them have similar initial locations and identical masses. Two of them are the only swarms that become the seeds for sweep-up, while the third is plotted for reference to show the evolution of an “average” particle in this model. This particle, after 25,000 yrs of evolution, clearly undergoes fragmentation.

Because of the smooth transition between sticking and bouncing, a limited number of particles manage to grow to the maximum size before they have drifted inwards. These particles have a chance to avoid the fragmenting collisions. This is because of two reasons. One of them is that the largest particles are drifting the fastest. What is more: the more massive the drifting particle is, the lower is the probability of fragmenting collision, due to the fact that the transition from fragmentation to mass transfer regime is at the projectile mass equal to 0.001 times the target mass.
Thus some “lucky” particles from the higher end of the mass distribution spectrum can reach the position, where the equal-size collisions are very unlikely, as most of the surrounding particles are more than the three order of magnitude in mass lower, and so collisions will primarily lead to sweep-up growth. In our model we observe two such seeds. 
 After $t = 27,500$~yrs, they have reached the pressure trap, so their drift is halted, and by the end of the simulation, they have reached m-sizes.

\begin{figure}
   \centering
   \includegraphics[width=\hsize]{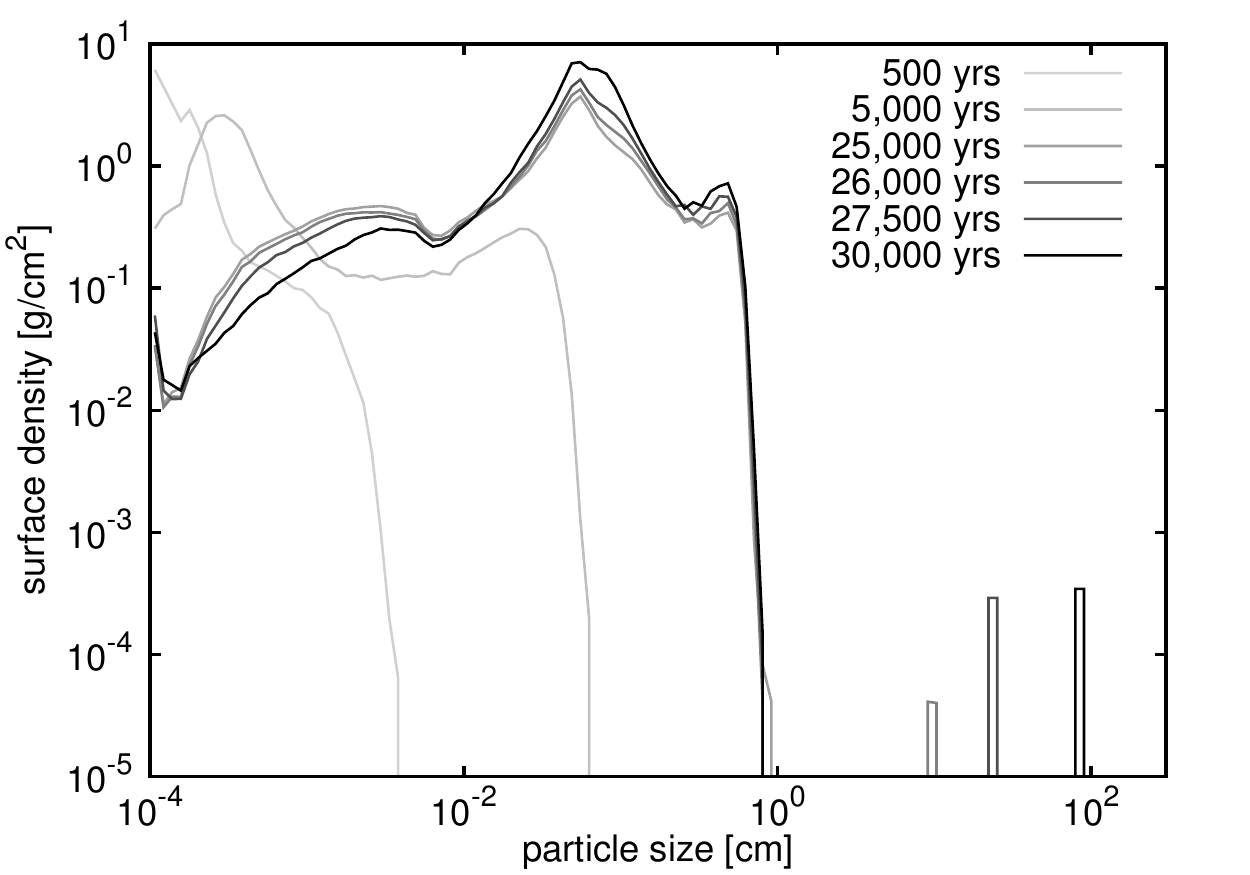}
      \caption{The evolution of the spatially integrated surface density of different sized particles for the model B. The particles grow until they reach the bouncing barrier. After that only a limited number of bodies continue the growth thanks to the mass transfer effect.}
         \label{fig:surfde_mt}
\end{figure}
\begin{figure}
   \centering
   \includegraphics[width=\hsize]{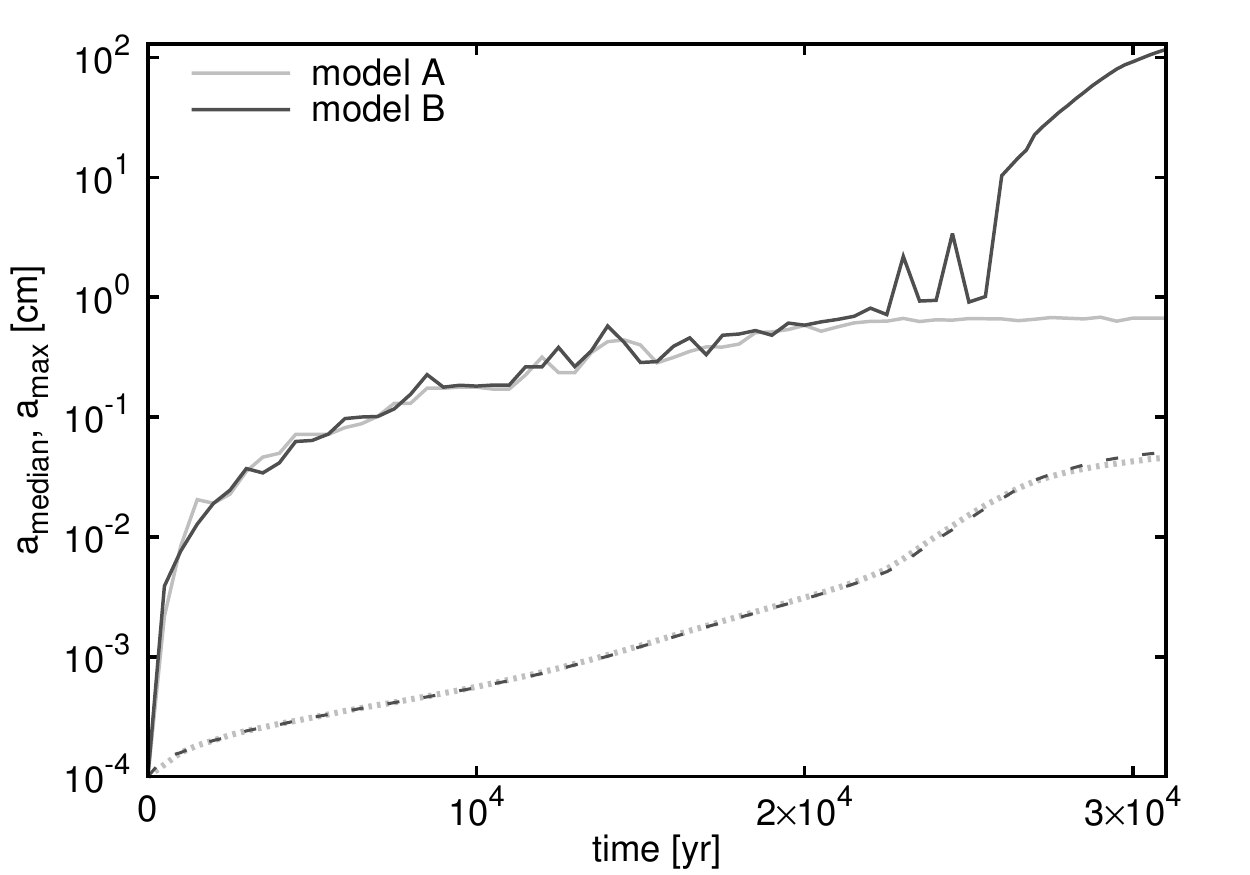}
      \caption{The time evolution the maximum (solid lines) and the median (dashed lines) radius of the dust particles in the model with and without the mass transfer effect. Both of the models generally evolve in similar way, what can be observed by following the median radius change. However, whereas the growth of the biggest particles is halted at sizes approximately $\sim$0.7 cm by the bouncing in the case of the model A, in model B some particles manage to grow till the radius of 100 cm after 30,000 yrs.}
         \label{fig:rmax}
\end{figure}

In Fig.\ \ref{fig:surfde_mt}, we present the evolution of the spatially integrated size-distribution for model B, and Fig.~\ref{fig:rmax} shows the evolution of the median and maximum radius of the dust particles for both models A and B. The median size illustrates the evolution of an “average” particle. Initially, both of the models evolve in the same way, and the median and maximum particle size grows gradually from $\mu$m- up to mm-sizes. After about $t = 23,000$~yrs, the largest particles in model A halt in their growth at a size of $0.7$ cm due to the bouncing barrier. The median size continues to grow, as all particles still have to reach the bouncing regime. In model B, however, the first seeds are formed and drifted into the pressure trap at $t = 23,000$~yrs, initiating sweep-up and causing a sudden increase in the size of the largest particles from cm- up to m-sizes. It can here be noticed that the sweep-up is only local, and the median particle size remains unchanged compared to model A. After the seeds are formed, the size-distribution becomes bimodal. As seen in Fig.\ \ref{fig:surfde_mt}, the seeds constitute a separate population, the surface density of which grows in time, as they are sweeping-up the small particles. However, as the seed population is represented by only two representative particles, its mass distribution cannot be resolved well in this model.

Fig.\ \ref{fig:rmax} shows also another interesting feature. The line corresponding to the maximum particle size in the model B exhibits two bumps before it finally jumps to depict the evolution of the two growing seeds: these are some “unlucky” particles that fragmented before being able to reach the region where the mass ratio between them and the small particles is high enough to lead to mass transfer collisions. Also later in our model we observe some particles that grow a bit bigger than the others but then fragment. Of course, there can and should be some new “lucky” seeds formed later, in particular from the particles that started their evolution further away in the disk. However, as the growth timescale rises with the distance from the star, we would have to carry on our models over longer time and possibly start with a larger domain to see more of the seeds, which is not possible with the current state of our code, as explained below.

The model B run took about 50 days on an 8 core 2.83 GHz Intel processor. The model A run took about half as long, as all the particles stayed small so that a longer advection time step was used. Because of the computational expense of the simulations, we finish them at $t = 31,000$~yrs, when the seeds have reached sizes of $100$~cm. Extrapolating the growth rate from the data, we find that the seeds would grow to km-sizes within the next $10^5$ yrs. \citet{2010ApJ...724.1153X} investigated the planetesimal growth by dust sweep-up and found the growth from 0.1 to 10 km within $10^6$ yrs in their models. However, in our model the surface density of the dust in the pressure trap is enhanced, as the solids are constantly delivered to the trap by the radial drift, so the timescale of the growth is reduced.

The dust density distribution in the dead zone exhibits a bump for particles smaller than 10$^{-2}$~cm (see Fig.\ \ref{fig:densar_mt}). This feature comes form the bimodal distribution revealed in the 1D tests presented in Sect.\ \ref{sub:1D}. It is not observed in the active zone while it is smeared by the relatively strong turbulence ($\alpha>10^{-4}$). Such structure cannot be modeled by the commonly used dust evolution codes that treat the vertical structure in an averaged way \citep{2010A&A...513A..79B}. However, the impact of resolving of this structure is hard to define without detailed comparison between results given by such 1D code and by our code, which is much beyond the scope of this work.

\section{Discussion and Conclusions}\label{sub:last}
We developed a new computational model for dust evolution in the protoplanetary disk. The representative particles method \citep{2008A&A...489..931Z} has been used to describe the dust. The gas disk has been included in an analytical way. The model tracks the dust drift as well as the coagulation. The Monte Carlo method has been used to investigate the collisions between the dust aggregates. The code is a further development of the work presented by \citet{2011A&A...534A..73Z}. We extended the model of \citet{2011A&A...534A..73Z} by adding an adaptive gridding method, which assures high spatial resolution in high dust density regions, as well as the radial dimension. 

With the new numerical code, we found that high spatial resolution is necessary to model the dust evolution properly. In particular, in the case of lack of turbulent mixing, when a dense midplane layer is formed, the dust growth timescale depends on the resolution very strongly. 

We noticed that a sharp change in a protoplanetary disk structure can be favorable for the planetesimal formation by sweep-up as suggested by \citet{2012A&A...540A..73W}. We applied our method to a snow line region in a low mass protoplanetary disk, and modeled the disk following the prescription of \citet{2007ApJ...664L..55K}, where the turbulent viscosity changes around the snow line, leading to occurrence of a low turbulence region as well as a pressure bump. We found that in such a disk it is indeed possible to grow planetesimals by the sweep-up. Due to the local dust density enhancement in the pressure bump, the sweep-up growth rate is increased, and the estimated planetesimal formation timescale is relevant for the planet formation.

The main conclusions of the models presented in this paper can be summarized in the following points.
\begin{itemize}
\item{Adaptive gridding allows us to investigate the dust collisional evolution with Monte Carlo method in 2D. It assures sufficient spatial as well as mass resolution to account for the dust structure. It automatically moves the computational effort towards the high dust density regions.}
\item{Proper resolution of the vertical structure of protoplanetary disk is important for obtaining a correct dust growth timescale. This is true especially in the case of low turbulence.}
\item{In some protoplanetary disks, it is likely to overcome the growth barriers and obtain planetesimals via the sweep-up growth, as suggested by \citet{2012A&A...540A..73W}. Since the disk properties change at different disk locations, the bouncing barrier is shifted in terms of maximum particles size. In our model the particles can grow more in the dead zone region. A limited number of these particles can become planetesimal seeds and continue to grow in the region of stronger turbulence, and their radial drift is at the same time halted because of the existence of the pressure trap.}
\item{The snow line, as modeled by \citet{2007ApJ...664L..55K}, is a favorable region for the planetesimal formation by incremental growth.}
\end{itemize}

Our model includes inevitable simplifications. We assumed that the disk is isothermal, i.e.\ the gas temperature $T_{\rm{g}}$ is constant along the vertical dimension. Also the turbulent viscosity $D_{\rm{g}}$ has been assumed to stay constant along a column at given distance from the star. We do not expect that including the dependence of $T_{\rm{g}}$ and $D_{\rm{g}}$ on the height above the midplane would influence the possibility of forming planetesimals via the mechanism described in this work.
We did not include the gas disk evolution. This is not consistent with the snow line model we implement, as the gas ionization rate and thus the turbulence strength is dependent on the dust properties. As we include the dust growth, the total surface area of grains changes and thus the turbulence is modified. The disk structure builds up on the disk accretion timescale. As the dust coagulation proceeds much faster than the gas disk evolution, we start our models from a steady state disk, which is a common, but not self-consistent practice.

We focused on the disk region around the snow line, where the temperature allows for existence of solid water ice. The collisional properties of ice aggregates are generally thought to be much better than silicates. However, due to lack of laboratory data, we did not include the ice in our models. We used a collision model that reflects the evolution of silicate grains. Including collision model of ices would help the growth of particles. On the other hand, it would also require considering another complex effects, such as evaporation and condensation \citep{2011ApJ...739...18K,2013A&A...552A.137R} or sintering \citep{2011ApJ...735..131S}. 

The planetesimal formation model we propose relies on the existence of a growth barrier, such as the bouncing barrier introduced by \citet{2010A&A...513A..57Z}. The robustness of the barrier has recently been put into question as the sticking and bouncing efficiency have been shown to exhibit a strong dependence on the internal structure of colliding aggregates as well as the impact parameter \citep{2011ApJ...737...36W, 2013arXiv1302.5532K}. Even though the results on the bouncing behavior is inconclusive, we argue that fragmentation could also work in a similar fashion for the sweep-up scenario. In case where fragmentation acts as the main growth barrier, the smaller dust population would be able to grow a bit further, but might still be swept up by the drifting seeds. As the mass transfer experiments have so far been performed over a limited parameter space only, this possibility would need to be verified experimentally.

As mentioned in Sect.\ \ref{sub:1D}, we did not include the effects of particle clumping via the streaming instability \citep{2005ApJ...620..459Y} and possibility of the gravitational instability \citep{2007Natur.448.1022J} for dense midplane layer. These effects could change our results and lead to efficient planetesimal formation in the dead zone. We plan to implement these phenomena in future work.

A great difficulty in the planet formation modeling is that the dynamic range involved is too wide to be covered by a single numerical method. When km-sized bodies are formed, regardless of their formation process, the gravitational interactions become important and the N-body dynamics needs to be considered. Statistical approach, which is commonly used to study the dust coagulation, is very hard to connect with the N-body methods, as it only handles the dust distribution function and it does not consider individual particles. In our code, we used the representative particles as a description for dust, thus the connection to the N-body regime is more natural. The code presented in this paper is a very first step towards a complete model of the planet formation.

We focused the study presented in this paper on the snow line region, which is shown to be a favorable region for formation of big bodies in protoplanetary disk. This is consistent with other studies. The impact of the snow line is reported in the context of the exoplanets distribution. \citet{2009ApJ...691.1322S} argue that the statistical properties of observed exoplanets cannot be explained without taking the snow line into account.
Also very recent work of \citet{2013MNRAS.428L..11M} suggests that the snow line region is where the asteroid belts are preferentially formed. They come to this conclusions because of a correlation between the location of the snow line and observed warm dust belts in exosolar systems. They argue that the existence of such asteroid belts may be crucial for existence of life on rocky planets.
It is worth noting that the planetesimal formation mechanism introduced in this work can take place also at locations other than the snow line.
\citet{2013ApJ...765..114D} have recently suggested that a steep variation of the turbulence efficiency and resulting pressure bump, which we need for our mechanism, can occur beyond what they call “metal freezeout line”, i.e. at the border beyond which metal atoms in the gas phase thermally adsorb on dust particles.
Anyway, in our model, the internal planetesimal population is formed at the same location in the disk that corresponds to the pressure maximum. Such a narrow annulus of planetesimals was suggested by \citet{2009ApJ...703.1131H} as initial condition for formation of the terrestrial planets in the Solar System.

The thresholds that change the structure of protoplanetary disk, such as the snow line, clearly have a great significance for the emergence and evolutions of planetary systems. Our models show that even with a “pessimistic” setup, in a low-mass protoplanetary disk consisting of not very sticky silicate grains, it is possible to form planetesimals at such a specific location. Further work is required to check the following evolution of the planetesimal ring. For bodies bigger than kilometers, consideration of the gravitational interactions, which we plan to include in our future work, is necessary. The question whether there will be an asteroid belt or a planet formed at the pressure trap cannot be answered at the current stage of our project.

\begin{acknowledgements}
We thank Carsten Dominik, Chris Ormel, Andras Zsom, Til Birnstiel and Paola Pinilla for useful discussions.
We thank our referee, Satoshi Okuzumi, for very quick and thorough report that helped us to substantially improve this paper.
J.D. was supported by the Innovation Fund FRONTIER of the Heidelberg University. F.W. was funded by the Deutsche Forschungsgemeinschaft within the Forschergruppe 759 “The Formation of Planets: The Critical First Growth Phase”.
J.D. would also like to acknowledge the use of the computing resources provided by bwGRiD (http:$\backslash\backslash$www.bw-grid.de), member of the German D-Grid initiative, funded by the Ministry for Education and Research (Bundesministerium für Bildung und Forschung) and the Ministry for Science, Research and Arts Baden-Wuerttemberg (Ministerium für Wissenschaft, Forschung und Kunst Baden-Württemberg)
\end{acknowledgements}

\bibliography{bib}
\end{document}